\documentclass[traditabstract]{aa} % for the abstract without structuration 
\usepackage{graphicx}
\usepackage{url}
\usepackage{lscape}
\usepackage{longtable}
\usepackage{multirow}
\usepackage{txfonts}
\newcommand{\hi}{H\textsc{i}}   %%%%%% nel testo $\mathrm{\hi}$
\begin{document}
   \title{Cold gas properties of the \textit{Herschel} Reference Survey. II. Molecular and total gas scaling relations}

   %\subtitle{Molecular and total gas properties of a volume-limited, K-band selected sample of nearby galaxies
   %\thanks{Based on observations collected at the Observatoire de Haute Provence (OHP) (France), operated by the CNRS}
   %}

  \author{A. Boselli\inst{1}
  	  ,
	  L. Cortese\inst{2,3}
	  ,
	  M. Boquien\inst{1}
	  ,
	  S. Boissier\inst{1}
	  ,
	  B. Catinella\inst{2,4}
	  ,
	  C. Lagos\inst{3}
	  ,
	  A. Saintonge\inst{5}
         }
	 
	 \authorrunning{Boselli et al.}
	 \titlerunning{Molecular and total gas scaling relations}
\institute{	
	Laboratoire d'Astrophysique de Marseille - LAM, Universit\'e d'Aix-Marseille \& CNRS, UMR7326, 38 rue F. Joliot-Curie, 13388 Marseille Cedex 13, France 
        \email{Alessandro.Boselli@lam.fr; mboquien@ast.cam.ac.uk;
	Samuel.Boissier@lam.fr}
	\and
	Centre for Astrophysics \& Supercomputing, Swinburne University of Technology, Mail H30, PO Box 218, Hawthorn, VIC 3122, Australia
	\email{lcortese@swin.edu.au; bcatinella@swin.edu.au}
	\and
	European Southern Observatory, Karl-Schwarzschild Str. 2, 85748 Garching bei Muenchen, Germany
        \email{clagos@eso.org}
	\and
	Max-Planck Institut f\"{u}r Astrophysik, D-85741 Garching, Germany
	\and
	Max-Planck-Institut fur Extraterrestrische Physik, 85741 Garching, Germany
	\email{amelie@mpe.mpg.de}
        }

   \date{}

% \abstract{}{}{}{}{} 
% 5 {} token are mandatory
 
  \abstract
  {We study the properties of the cold gas component of the interstellar medium of the \textit{Herschel} Reference Survey, a complete volume-limited (15$\lesssim$ $D$ $\lesssim$ 25 Mpc), 
  K-band-selected sample of galaxies spanning a wide range in morphological type (from ellipticals to dwarf irregulars) and stellar mass (10$^9$ $\lesssim$ $M_{star}$ $\lesssim$ 10$^{11}$ M$_{\odot}$). 
  The multifrequency data in our hands are used to trace the molecular gas mass distribution and the main scaling relations of the sample, which put strong constraints
  on galaxy formation simulations. We extend the main scaling relations concerning the total and the molecular gas component
  determined for massive galaxies ($M_{star}$ $\gtrsim$ 10$^{10}$ M$_{\odot}$)
  from the COLD GASS survey down to stellar masses $M_{star}$ $\simeq$ 10$^{9}$ M$_{\odot}$. As scaling variables we use the total stellar mass $M_{star}$, the stellar surface density $\mu_{star}$, 
  the specific star formation rate SSFR, and the metallicity of the target galaxies. By comparing molecular gas masses determined using a constant or a luminosity dependent $X_{CO}$ conversion
  factor, we estimate the robustness of these scaling relations on the very uncertain assumptions used to transform CO line intensities into molecular gas masses. 
  The molecular gas distribution 
  of a K-band-selected sample is significantly different from that of a far-infrared-selected sample 
  since it includes a significantly smaller number of objects with $M(H_2)$ $\lesssim$ 6 10$^9$ M$_{\odot}$.
  In spiral galaxies the molecular gas phase 
  is only 25-30\% ~ of the atomic gas. The analysis also indicates that the slope of the main 
  scaling relations depends on the adopted conversion factor. Among the sampled relations, all those concerning $M(gas)/M_{star}$ are 
  statistically significant and show little variation with $X_{CO}$. We observe a significant correlation between $M(H_2)/M_{star}$ and $SSFR$, $M(H_2)/M(\mathrm{\hi})$ and 
  $\mu_{star}$, $M(H_2)/M(\mathrm{\hi})$, and 12+log(O/H) regardless of the adopted $X_{CO}$. 
  The total and molecular gas consumption timescales are anticorrelated with the specific star formation rate. 
  The comparison of HRS and COLD GASS data indicates that some of the observed scaling relations are nonlinear.
  }
   {}
   {}
   {}
   {}
   {}

	       
   \maketitle
%
%________________________________________________________________

\section{Introduction}

The cold gas phase of the interstellar medium (ISM) is one of the main ingredients in the process of star formation. The atomic gas component formed after
the Big Bang collapses within dark matter haloes to form molecular clouds. Stars are formed within these molecular clouds. The same stars 
produce and inject metals into the ISM, part of them aggregating to form dust grains. The dust component plays a major role regulating the star
formation process within galaxies. It first acts as a catalyst in the process that transforms the atomic hydrogen into its molecular phase by absorbing and scattering the
ionising and non-ionising light of the interstellar radiation field, it participates in the energetic balance that regulates the heating and cooling process of the ISM.
Acting as a screen, it also prevents the dissociation of molecules.\\ 

This matter cycle is critical for the study of the formation and evolution of galaxies (e.g. Boselli 2011). It can be
studied by either locally analysing the relationships between the different properties of the ISM on kpc scales (gas, dust, metals, radiation field) and its 
relations with the star formation process or by using integrated values of statistically significant samples of galaxies spanning a wide range in morphological type and
luminosity. The first approach has been recently developed by the THINGS team. Thanks to the high spatial resolution of the multifrequency data available to them, Leroy et al.
(2008) and Bigiel et al. (2008; 2011) have studied the relationships between the star formation activity and the atomic and molecular gas
content of a sample of $\sim$ 20-30 galaxies on sub-kpc scales in the nearby universe. The results that they obtained allowed the community to make a major step towards understanding the
process of star formation in galaxies. They confirmed the existence of a tight relation between the gas and the star-formation-rate surface density, generally called Schmidt
law (Schmidt 1959; Kennicutt 1998), thereby extending earlier results from Kennicutt (1989), Martin \& Kennicutt (2001), and Wong \& Blitz (2002)
on a local scale of a few 100 pc. They have also observationally clearly
confirmed that the gas phase that regulates the star formation process is mainly the molecular one (Bigiel et al. 2008), as is indeed expected from theory. 
The gas has to cool and collapse to form molecular clouds where star formation takes place; see however Krumholz (2012). At the same time, they have also shown that
the efficiency of transforming gas into stars depends on several physical parameters such as the stellar and gaseous density, and the midplane pressure or the orbital timescale
(Blitz \& Rosolowsky 2006; Leroy et al. 2008; Kennicutt \& Evans 2012; Momose et al. 2013). \\

Because of the small sample and the limited range covered in parameter space, however, it soon became clear that these
works needed to be extended to be representative of the whole galaxy population. The same THINGS team has recently made several efforts to extend their analysis to different
physical conditions, such as those encountered in the outskirt of extended discs (Bigiel et al. 2010), in H{\sc i}-dominated regions (Schruba et al. 2012), or in metal-poor regions (Bolatto et al. 2011). 
Other attempts have been made to study the star formation process in cluster galaxies (Fumagalli \& Gavazzi 2008; Vollmer et al. 2012), where the interaction with the hostile environment 
can perturb this process (Boselli \& Gavazzi 2006). The star formation process has also been studied in other extreme environments, such as in
the ram pressure stripped gas ejected into the hot intracluster medium (Boissier et al. 2012). \\

A different approach is the one adopted by the GASS/COLD GASS team, who defined and observed a statistically significant sample of nearby massive galaxies in the H{\sc i} and CO lines 
($M_{star}$ $\gtrsim$ 10$^{10}$ M$_{\odot}$) extracted from the SDSS
with available UV \textit{GALEX} data necessary to quantify the star formation activity (Catinella et al. 2010, 2012; Saintonge et al 2011a, 2011b).
By addressing questions similar to those already discussed in previous works (Young \& Knezek 1989; Young \& Scoville 1991; Sage 1993; Boselli et al. 1995, 1997, 2002; Young et al. 1996; 
Casoli et al. 1998; Sauty et al. 2003, and more recently Lisenfeld et al. 2011), the GASS/COLD GASS team has awoken new interest in the study of the cold gas component by means of the scaling relations.
The importance of scaling relations resides in the fact that they are generally used as strong observational constraints in models of 
galaxy formation and evolution. At the same time, they can be used to study the relationships between the different galaxy constituents (gas, dust, metals,
stellar populations, dark matter, etc.) and are thus powerful tools for studying the different galaxy populations. 
These works soon became a reference for cosmological studies based on semi-analytic models of galaxy evolution, now able to trace the evolution of the different 
gas phases (Gnedin et al. 2009; Dutton 2009; Dutton \& van den Bosch 2009; Fu et al. 2010; Power et al. 2010; Cook et al. 2010; Lagos et al. 2011a,b; Kauffmann et al. 2012). 
In addition to defining new scaling relations relative to the molecular gas component, the study of the
COLD GASS sample has shown that the molecular gas depletion timescale is not
constant but instead changes in galaxies (Saintonge et al. 2011b). This result is surprising since it apparently contradicts
the local studies of Bigiel et al. (2008) and Leroy et al. (2008) (see however Momose et al. 2013).\\

With the aim of studying the star formation process in detail and more generally the matter cycle in galaxies of different morphological type and stellar mass, 
we undertook a multifrequency survey covering the electromagnetic spectrum from the UV to the radio centimetric of a well-defined,  K-band-selected, 
volume-limited sample of nearby galaxies: the \textit{Herschel} Reference Survey (HRS; Boselli et al. 2010a). 
This sample includes more than three hundred objects and is ideally defined to characterise the statistical properties of normal, nearby galaxies. Given the tight
relation between the near-infrared bands and the total stellar mass of galaxies (Gavazzi et al. 1996), the choice of the K-band secures a stellar mass selection.  
The sample, which includes both isolated galaxies and objects in the Virgo cluster, 
is also appropriate for studying the effects of the environment on galaxy evolution (Boselli \& Gavazzi 2006).\\

The HRS has been recently observed in guaranteed time (Ciesla et al. 2012) with the 
SPIRE instrument (Griffin et al. 2010) on board \textit{Herschel} (Pilbrat et al. 2010).
We have also gathered H{\sc i} and CO data from our own observations or from the literature necessary for constraining the gas content (Boselli et al. 2013a), 
narrow band H$\alpha$ imaging tracing the ionising radiation produced by newly formed massive stars (Boselli et al. in prep), \textit{GALEX} FUV and NUV data 
to estimate the star formation activity and the dust attenuation (combined with infrared data; Boselli et al. 2011; 
Cortese et al. 2012a). Integrated optical spectroscopy, necessary for
measuring the mean gas metallicity, is also available for the late-type systems (Boselli et al. 2013b; Hughes et al. 2013). This sample has 
been already used to trace the typical scaling relations relative to the atomic gas content (Cortese et al. 2011) and of the dust mass (Cortese et al.
2012b), as well as for studying the physical properties of the different dust components (Boselli et al. 2010b, 2012). \\

The main purposes of the present paper are to extend the scaling relations determined for the COLD GASS 
sample down to stellar masses $M_{star}$ $\simeq$ 10$^{9}$ M$_{\odot}$ and to quantify the robustness of these relations vs. the use of a 
constant or a variable $X_{CO}$ conversion factor. The important results obtained so far by the COLD GASS team are based on a sample
composed only of massive galaxies ($M_{star}$ $\gtrsim$ 10$^{10}$ M$_{\odot}$), so cannot be taken as representative of the entire galaxy population. 
The HRS is ideally defined for such a purpose since it is a statistically complete, volume-limited sample that extends the dynamic range of the COLD GASS survey
by $\sim$ a factor of ten in stellar mass.
Another novelty with respect to the COLD GASS survey is that we can trace, for the first time in the literature, 
the main scaling relations of the total and molecular gas components as a function of the metallicity, one of the key parameters characterising
the physical properties of the interstellar medium.\\

Dominated by massive objects, molecular gas masses of COLD GASS galaxies have been determined by assuming a constant conversion factor.
In recent years it has become evident that the $X_{CO}$ conversion factor 
is not universal, but instead changes with the physical properties of the ISM (e.g. Kennicutt \& Evans 2012 and the recent review of Bolatto et al. 2013). 
Given the wide dynamic range in the parameter space covered by the HRS galaxies (Boselli et al. 2012), we can expect that within our sample the $X_{CO}$ conversion
factor changes significantly from massive, metal-rich quiescent galaxies to dwarf, metal-poor objects  
characterised by a harsh interstellar radiation field. We thus test the robustness and the universality of
these newly determined scaling relations by using either a constant or a variable CO-to-H$_2$ conversion factor. \\

The present work is based on the recent compilation of atomic and molecular
gas data for the HRS galaxies presented in a companion paper, Boselli et al. (2013a; paper I). 
The paper is organised as follows: In Sects. 2 and 3 we describe the sample and the physical variables obtained from ancillary data necessary for the analysis. 
In Sect. 4 we reconstruct the molecular gas mass distribution of the HRS sample. In  Sect. 5 we describe how the molecular and total gas content of galaxies change with morphological type.
In Sects. 6 and 7 we determine the major scaling relations defined by the molecular and total gas content and the
depletion time and then discuss the implications of these relations. 
As structured, the present paper revisits the main topics addressed in our previous work (Boselli et al. 2002) on a much more homogeneous, complete, and high-quality data set. 
In a third paper (Boselli et al. 2013c; paper III of this series), we compare the mean properties of HRS field galaxies to those of Virgo cluster members 
to study whether the hostile cluster environment is able to perturb the molecular gas phase of late-type galaxies (Boselli \& Gavazzi 2006).\\

\section{The sample}

\begin{figure}
   \centering
   \includegraphics[width=9cm]{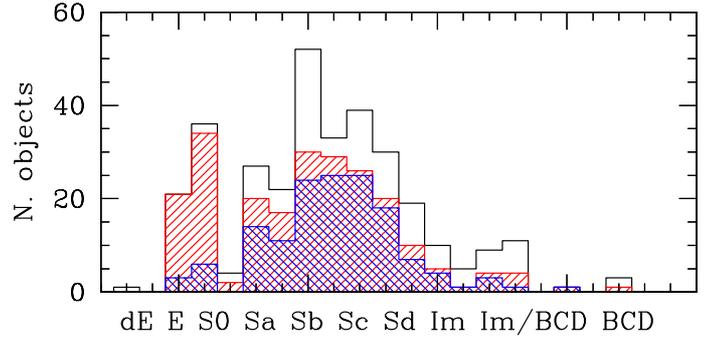}
   \caption{Distribution in morphological type of all the HRS galaxies (black solid line) and of those objects with available CO data (red shaded
   histogram). Galaxies detected in CO are represented by the blue shaded histogram. }
   \label{type}%
   \end{figure}

The \textit{Herschel} Reference Survey is a SPIRE guaranteed time key project aimed at observing with \textit{Herschel} a complete, 
K-band selected (K $\leq$ 8.7 mag for early-types, K $\leq$ 12 mag for type $\geq$ Sa), volume-limited (15$\leq$ $D$ $\leq$ 25 Mpc)
sample of nearby galaxies at high galactic latitude. The sample, which is extensively presented in Boselli et al. (2010a),
is composed of 322 galaxies out of which 260 are late-type systems\footnote{With respect to the original sample given in Boselli et al. (2010a), we 
removed the galaxy HRS 228 whose new redshift indicates it as a background object.
We also revised the morphological type for 6 galaxies: 
NGC 5701, now classified as Sa, NGC 4438, and NGC 4457 now Sb, NGC 4179, now S0, VCC 1549, now dE, and NGC 4691 now Sa.}. 
Figure \ref{type} shows the distribution of the different morphological types within the HRS.
The K-band selection has been chosen as a proxy for galaxy stellar mass (Gavazzi et al. 1996). The sample includes objects in environments of 
different density, from the core of the Virgo cluster, to loose groups and fairly isolated systems. As defined, the present
sample is ideal for any statistical study of the mean galaxy population of the nearby universe.\\

\begin{figure}
   \centering
   \includegraphics[width=9cm]{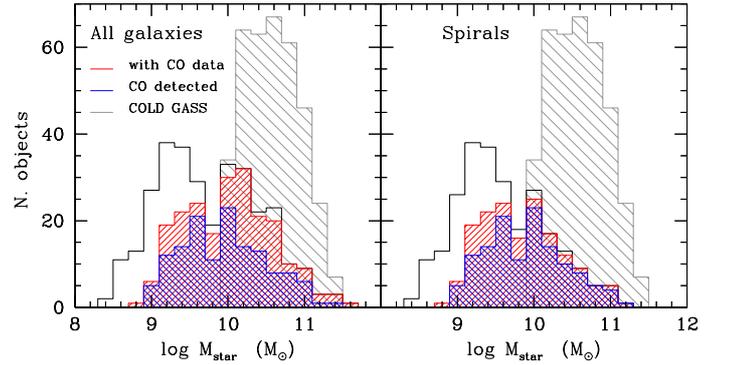}
   \caption{Distribution in stellar mass of the HRS galaxies (black solid line) and of those objects with available CO data (red shaded
   histogram) for the whole sample (left) and for late-type objects (right). 
   Galaxies detected in CO are represented by the blue shaded histogram. The grey shaded histogram shows the 
   distribution in stellar mass of the COLD GASS galaxies with CO data.}
   \label{completezza}%
   \end{figure}

In this paper we study the gas properties of this volume-limited, complete sample of nearby galaxies. H{\sc i} data are available for almost the entire sample (98\%).
The present work will thus be limited to the subsample of galaxies with an available estimate of the molecular gas content (see next section).
At present CO data are available for 225 out of the 322 galaxies of the sample (see Table \ref{Tabdetection}). Figure \ref{type} shows that the sample is almost complete
for early-type galaxies, while it is still incomplete for late-type objects. When limited to a K band magnitude of K$\leq$ 10 mag, 94\% ~ 
of the HRS late-type systems have molecular gas data (Boselli et al. 2013a). Figure \ref{type} also shows that all morphological classes 
from Sa to Im and BCD are represented in the subsample of the HRS with available CO data. The detection rate is very low for early types
(16 \%) and good for late types (80\%). Figure \ref{completezza} shows the distribution in stellar mass of the HRS galaxies 
with CO data. CO data are available for galaxies spanning the range in stellar mass 
10$^9$ $\lesssim$ $M_{star}$ $\lesssim$ 3 10$^{11}$ M$_{\odot}$, and the sample is almost complete for $M_{star}$ $\gtrsim$ 3 10$^9$ M$_{\odot}$.
The HRS sample analysed in this work can thus be considered as representative 
of the nearby universe down to this limit. This dynamic range is significantly wider than the one covered by the COLD GASS survey (Saintonge et al. 2011a), which is
limited to $M_{star}$ $>$ 10$^{10}$ M$_{\odot}$.

\begin{table*}
\caption{HRS galaxies with CO and H{\sc i} data}
\label{Tabdetection}
{%\scriptsize
\[
\begin{tabular}{cccccc}
\hline
\hline
\noalign{\smallskip}
\hline
Type	        & total &observed CO	& detected CO   & Observed H{\sc i}  & detected H{\sc i} \\
\hline
All       	& 322 	& 225       	& 143		& 315		& 244	\\
E-S0a           & 62    & 57		& 9		& 60		& 19    \\
Sa-Im-BCD       & 260	& 168		& 134		& 255		& 243   \\  
\noalign{\smallskip}
\hline
\end{tabular}
\]
}
\end{table*}

\section{The data}

\subsection{The gas component}

Integrated H{\sc i} data are available for almost all the late-type galaxies of the sample. They have been collected from the literature and have
been homogenised as described in Boselli et al. (2013a). H{\sc i} masses have uncertainties of the order of 15\%. 
H{\sc i} data are also used to quantify the effects induced by the Virgo cluster environment
on the gas content of the HRS galaxies. This is done through the H{\sc i}-deficiency parameter ($\mathrm{\hi}-def$), defined as the difference 
in logarithmic scale between the expected and the observed H{\sc i} mass of a galaxy  
(Haynes \& Giovanelli 1984). The expected atomic gas mass is the mean H{\sc i} mass of a galaxy of a given optical size and morphological type 
determined in a complete sample of isolated galaxies taken as reference. The H{\sc i}-deficiency parameter is poorly constrained in early-type galaxies. %We thus do not consider it in the following analysis.
H{\sc i} deficiencies for all the target galaxies have been measured using the recent calibrations of Boselli \& Gavazzi (2009).
We consider those late-type galaxies with an H{\sc i}-deficiency parameter $\mathrm{\hi}-def$ $\leq$ 0.4\footnote{The choice of a threshold in the H{\sc i}-deficiency
parameter $\mathrm{\hi}-def$ $\leq$ 0.4 to identify unperturbed objects is quite arbitrary. Different values have been used in other works, ranging from 0.3 (Gavazzi et al. 2013) to 0.5 (Cortese et al. 2011,
2012b). The main results of this work do not change if this threshold is kept within the 0.3 $\leq$ $\mathrm{\hi}-def$ $\leq$ 0.5.} as unperturbed objects.
Since the principal aim of this work is to determine the main relations that are valid for normal, unperturbed late-type galaxies, the fit to the data, unless specified, 
is always done only on the subsample of objects with a normal atomic gas content ($\mathrm{\hi}-def$ $<$ 0.4; filled symbols in the following figures). 
Galaxies with a larger H{\sc i}-deficiency are plotted here for completeness. \\
%They are considered in Boselli et al. (2013c) in the study of the effects of the environment on the cold gas properties of Virgo cluster galaxies.\\

Molecular hydrogen masses of the HRS galaxies are determined from the $^{12}$CO(1-0) spectroscopic data recently collected and homogenised in Boselli et
al. (2013a). CO intensities can be converted into molecular hydrogen masses using a CO-to-H$_2$ conversion factor, $X_{CO}$.  
As extensively discussed in the literature, however, this conversion factor is expected to vary with the physical properties of the interstellar medium of the observed galaxies 
(interstellar radiation field, metallicity, etc.; e.g. Bolatto et al. 2013). The exact dependence of $X_{CO}$ on these parameters
is, however, still strongly debated. In fact, observations and models still give inconsistent results (Boselli et al. 2002; Bolatto et al. 2008; Shetty et al. 2011a; Schruba et al. 2012; 
Kennicutt \& Evans 2012; Bolatto et al. 2013; Sandstrom et al. 2013). Since with the present set of data we are not able
to test whether $X_{CO}$ is constant or variable, we thus decided to determine the molecular hydrogen
mass in two different ways. In the first one we adopt a constant conversion factor of $X_{CO}$ = 2.3 10$^{20}$ cm$^{-2}$/(K km s$^{-1}$) 
($\alpha_{CO}$ = 3.6 M$_{\odot}$/(K km s$^{-1}$ pc$^2$)) of Strong et al. (1988), and
in the second one we use the H-band luminosity-dependent $X_{CO}$ of Boselli et al. (2002)

\begin{equation}
{log X_{CO} =  -0.38 \times log L_H + 24.23  ~~~~\rm{[cm^{-2}/(K km s^{-1})]}}
\end{equation}

H-band luminosities are available from 2MASS for all the HRS galaxies. This calibration gives values of $X_{CO}$ ranging from $X_{CO}$ = 3.9 10$^{20}$ cm$^{-2}$/(K km s$^{-1}$)
in low-luminosity systems to $X_{CO}$ = 8.6 10$^{19}$ cm$^{-2}$/(K km s$^{-1}$) in massive spirals. These values of $X_{CO}$ can be compared to those of Obreschkow \& Rawlings (2009)
determined for galaxies with similar CO luminosities (7.5 10$^{19}$ $\leq$ $X_{CO}$ $\leq$ 2.7 10$^{20}$ cm$^{-2}$/(K km s$^{-1}$)) or to those predicted by
the models of Feldmann et al. (2012) (1.6 10$^{20}$ $\leq$ $X_{CO}$ $\leq$ 5 10$^{20}$ cm$^{-2}$/(K km s$^{-1}$)) and Narayanan et al. (2012) (10$^{20}$ $\leq$ $X_{CO}$ $\leq$ 2.5 10$^{20}$ cm$^{-2}$/(K km s$^{-1}$))
for galaxies in the same metallicity range as those analysed in this work. 
We also notice that this calibration in the 
CO-to-H$_2$ conversion factor is perfectly consistent with the most recent results obtained with \textit{Herschel} on nearby galaxies by 
Sandstrom et al. (2013). In this way we can study how robust are the inferred scaling relations with respect to the poorly constrained conversion factors. 

Beyond the evident uncertainty introduced by the use of a constant or a luminosity dependent conversion
factor, the molecular gas determination is also strongly affected by the uncertainty due to the extrapolation of single beam observations to total
fluxes. For the HRS galaxies, this source of uncertainty has been quantified in Boselli et al. (2013a). It generally goes from $\sim$ 
12\% ~ in galaxies fully mapped in CO from Kuno et al. (2007), to $\sim$ 44\% ~ in other mapped galaxies or in objects where 
the single beam observations cover more than 10\% ~ of the surface of the galaxies, to $\sim$ 100\% ~ in the remaining objects. 
Since the HRS sample is dominated by galaxies with single beam observations covering more than 10\% ~ of the total surface of the galaxy, we assume that the mean uncertainty on 
the total CO flux of our targets is $\sim$ 44 \%. 
\\ 

Total gas masses have been determined by combining the molecular and atomic gas content described above, and including an helium contribution of 30
\%

\begin{equation}
{M(gas) = 1.3 [M(\mathrm{\hi})+M(H_2)] ~~~~~~~~\rm{[M_{\odot}]}}
\end{equation}

\noindent
The constant conversion factor used to infer the total molecular gas mass of the HRS galaxies has been determined using gamma-ray data observed by \textit{COS-B} (Strong et al. 1988), 
and thus includes the contribution of the dark molecular gas (Wolfire et al. 2010). The luminosity-dependent conversion factor relation adopted in this work has been 
calibrated using a different set of data, from $\gamma$-ray data, virial mass estimates of giant molecular clouds, metallicity dependent dust-to-gas ratios, and thus does not necessarily
include the contribution of the dark molecular hydrogen (Wolfire et al. 2010). This luminosity-dependent calibration, however, fits the Milky Way $\gamma$-ray determination well, 
and should thus be fairly representative of the whole molecular hydrogen content. The recent models of Wolfire et al. (2010) indicate that the fraction of dark molecular hydrogen with respect
to that traced by the CO line is only $\simeq$ 30\%. The uncertainty in the total molecular hydrogen mass in our sample due to a possible inaccurate estimate of the dark component 
is thus smaller than the uncertainty due to the extrapolation of the observed CO flux for determining the total CO luminosity of the target galaxies.
We can also add that in normal galaxies such as those sampled in this work, the mean gas-to-dust ratio is $\sim$ 160 (Sodroski et al. 1994)
and slightly changes with metallicity (Boselli et al. 2002). The total gas mass 
determined using Eq. 1 can thus be taken as representative of the total cold baryonic mass of galaxies.\\

\subsection{Scale variables}

Stellar masses are derived from $i$-band luminosities using the $g-i$ 
colour-dependent stellar mass-to-light ratio relation from Zibetti et al. (2009), assuming a Chabrier (2003) initial mass function (IMF). This recipe 
for determining the stellar mass has been chosen for consistency with the previous works devoted to the study of the multifrequency scaling
relations of the HRS galaxies (Cortese et al. 2011, 2012a, 2012b) and is slightly different from the one generally used in other multifrequency studies done
by our team (e.g. Boselli et al. 2009; 2012). The direct comparison of the stellar masses determined using the two different recipes, however, indicates that
they are perfectly consistent\footnote{The ratio in log scale of the stellar masses determined as in Boselli et al. (2009) to the one measured in this work
is 0.99 $\pm$ 0.02.}. We assume that the typical uncertainty in the stellar mass estimate is 0.15 dex. Stellar mass 
surface densities, in M$_{\odot}$ kpc$^{-2}$, are determined through the relation

\begin{equation}
{\mu_{star} = \frac{M_{star}}{2 \pi R_{50,i}^2}  ~~~~~~~~ \rm{[M_{\odot} kpc^{-2}]}}
\end{equation}

\noindent
where $R_{50,i}$ is the radius containing 50\% of the $i$-band light, i.e. the effective radius (from Cortese et al. 2012b).\\
The specific star formation rate is defined as (Brinchmann et al. 2004)

\begin{equation}
{SSFR = \frac{SFR }{M_{star}} = \frac{b}{t_0 (1-R)} ~~~~~~~~  \rm{[yr^{-1}]}}
\end{equation}

\noindent
where $SFR$ is the star formation rate (in M$_{\odot}$ yr$^{-1}$) and $t_0$ the age of the galaxies. If we consider $R$ the returned gas fraction, generally taken equal to $R$=0.3 
(Boselli et al. 2001), the specific SFR can be converted into the birthrate parameter $b$ (Sandage 1986).
Thanks to the available multifrequency data in our hand, the birthrate parameter $b$ and the SSFR can be determined using different and independent sets
of data. Under some assumptions on the stationarity of the star formation activity of the target galaxies, UV, far infrared, and Balmer line emission data
can be transformed into SFRs using simple relations (see Boselli et al. 2009 for a critical discussion of these assumptions).
Here the SFR is determined using \textit{WISE} 22 $\mu$m data, available for all the HRS galaxies (Ciesla et al. in prep.),
combined with \textit{GALEX} FUV data (Cortese et al. 2012a) and using the calibration of Hao et al. (2011) as described in Cortese (2012). 

This calibration was originally defined for
the \textit{Spitzer} MIPS 24 $\mu$m band, but given the similarity of the flux densities in the two adjacent bands (\textit{WISE} 22 $\mu$m and MIPS 24 $\mu$m;
Ciesla et al. in prep.), it can be
adopted without any correction. The calibration of the SSFR is significantly different from the one we used in our previous works (Boselli et al.
2010b; Boselli et al. 2012) based on H$\alpha$+[NII] and \textit{GALEX} FUV imaging data, with the former corrected for dust extinction using the Balmer decrement and
for [NII] contamination using the integrated spectroscopic data of Boselli et al. (2013b), the latter using the extinction correction recipes of Cortese
et al. (2008) based on the far infrared-to-UV flux ratio (see Boselli et al. 2009). The two independent measurements of $SSFR$ are perfectly consistent\footnote{The ratio 
in log scale of the $SSFR$ determined as in Boselli et al. (2009, 2012) to that measured in this work
is 0.99 $\pm$ 0.04.}. The mean uncertainty on this variable is 0.25 dex as determined from the uncertainties on the stellar mass and
on the SFR of the target galaxies. 
The use of the \textit{WISE} and \textit{GALEX} data has been chosen because both sets of data are available for the whole sample. 
As defined, $SFR$ and $SSFR$ thus correctly account for dust
attenuation effects. \\

The present set of data is also used to estimate the gas depletion timescale or the star formation efficiency, defined as (Young et al. 1996; Boselli et al. 2001, 2002)

\begin{equation}
{\tau_{H_2} = \frac{M(H_2)}{SFR}  ~~  \rm{[yr]} = \frac{1}{SFE_{H_2}} ~~  \rm{[yr^{-1}]}}
\end{equation}

\noindent
when only the molecular gas content is considered, or equivalently

\begin{equation}
{\tau_{gas} = \frac{M(gas)}{SFR}    ~~\rm{[yr]} = \frac{1}{SFE_{gas}} ~~  \rm{[yr^{-1}]}}
\end{equation}

\noindent
when the total gas content is taken into account. As defined, the gas depletion timescale corresponds to the Roberts time 
if the recycled gas fraction is also considered (Roberts 1963; Boselli et al. 2001). We recall that these gas depletion timescales are
just a rough estimate of the time that galaxies can still sustain star formation. Because of the tight correlation between 
the star formation activity and the gas column density known as the Schmidt law (Schmidt 1959, Kennicutt 1998), in a closed box scenario, the 
decrease in the gas column density consumed via the formation of new stars induces a decrease in the star formation activity itself, making the
timescale for gas consumption longer than the one deduced from the previous relations. Furthermore, in a more realistic scenario, 
the infall of fresh gas that is expected in unperturbed, field galaxies should also be considered. \\

Metallicities are determined using the integrated spectroscopy of Boselli et al. (2013b) as indicated in Hughes et al. (2013). This is done using different sets of
optical emission lines and adopting as base metallicity the PP04 O3N2 calibration on [NII] and [OIII] emission lines (Pettini \& Pagel 2004). The different
12+log(O/H) estimates are then converted to the reference metallicity using the prescriptions of Kewley \& Ellison (2008), and averaged. The mean uncertainty relative to the adopted method on the
12+log(O/H) index is 0.13 dex (Hughes et al. 2013). \\
All multifrequency data used in this analysis have been published in dedicated works, and are made available to the community through our dedicated database
HeDaM (http://hedam.lam.fr/HRS/).

\subsection{COLD GASS data}

Throughout the paper we compare the HRS data to those of the COLD GASS sample taken from Saintonge et al. (2011a). The H{\sc i} data are updated with the final release of Catinella et al. (2013).  
For consistency, the COLD GASS molecular gas masses are corrected by a factor of 1.11\footnote{The conversion factor used in COLD GASS 
is $X_{CO}$ = 2.0 10$^{20}$ cm$^{-2}$/(K km s$^{-1}$).} whenever a constant $X_{CO}$ conversion factor is used in the analysis or when using 
the same H-band luminosity-dependent conversion factor adopted for the HRS galaxies\footnote{H-band luminosities of COLD GASS galaxies are determined from stellar masses using the relation
\begin{equation}
{log L_H = 0.83(\pm0.02) \times log M_{star} + 2.03(\pm0.15)}
\end{equation}
\noindent
determined in this work on the HRS sample, where stellar masses and H-band luminosities are expressed in solar units.}.
Metallicities are not available for the COLD GASS sample. For comparison with the HRS, we calculate the metallicity of these galaxies using the mass-metallicity relation
determined by Kewley \& Ellison (2008) using the PP04 O3N2 calibration. This is a row estimate of 12$+$log(O/H) and should be considered with caution in analysing
and interpreting of the scaling relations traced by COLD GASS galaxies.

\noindent
Although plotted in the figures, we expressly exclude COLD GASS galaxies from the determination of the mean gas scaling relations for two main reasons: \\
1) we want to determine these relations in a K-band selected, volume-limited sample of galaxies (the HRS); and\\
2) we want to use the most homogeneous set of data. With respect to this second point, the data of the COLD GASS sample differ from ours in the determination
of the total CO flux, the star formation activity, and of the metallicity. In COLD GASS the total CO flux of the observed galaxies is
extrapolated using the empirical relations presented in Saintonge et al. (2011a), while for the HRS objects we either use CO maps or we extrapolate single beam
observations assuming a 3D exponential CO distribution (Boselli et al. 2013a). While the $SFRs$ of the HRS galaxies are directly measured from the 
\textit{GALEX} FUV and \textit{WISE} mid-infrared data,
%and are thus properly taking into account the effects due to dust attenuation
those of the COLD GASS sample are determined using the UV-optical SED fitting technique (Saintonge et al. 2011b).

\section{The molecular gas mass distribution}

\begin{figure}
   \centering
   \includegraphics[width=9cm]{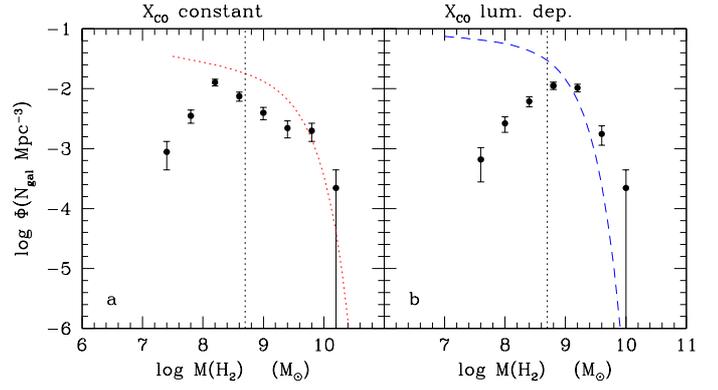}
   \caption{The molecular hydrogen mass distribution of the HRS galaxies in bins of 0.4 in log.
   The molecular gas mass is determined assuming a constant $X_{CO}$ factor (left panel) or the luminosity dependent $X_{CO}$ factor 
   given in Boselli et al. (2002) (right panel). 
   The distribution is compared to the one derived by Oberschkow \& Rawlings (2009) from the CO luminosity distribution 
   of Keres et al. (2003) assuming a constant (red dotted line) or a luminosity dependent (blue dashed line) $X_{CO}$ conversion
   factor. The vertical black dotted line indicates the completeness in $M(H_2)$ of the HRS sample.
   }
   \label{H2MF}%
   \end{figure}

Determination of the molecular gas mass function from the HRS galaxies, which is comparable to the H{\sc i} mass function determined by Zwaan et al.
(2005) using HIPASS data and Martin et al. (2010) with ALFALFA data, is impossible since our set of data is not a sample of galaxies extracted from a blind CO survey 
of a representative volume of the universe. Because of its selection, we can only determine the molecular gas mass distribution for 
a K-band-selected, volume-limited sample of nearby galaxies.
Figure \ref{H2MF} shows the distribution of the molecular gas mass within the HRS subsample of late-type galaxies with $\mathrm{\hi}-def$ $\leq$ 0.4. The number of
galaxies per bin of molecular gas mass of width 0.4 dex is computed including upper limits but counting them in the nearest
lower mass bin. %To use a complete sample, we limit the determination of the molecular gas distribution to galaxies with stellar
%mass $M_{star}$ $\geq$ 3 10$^9$ M$_{\odot}$. Within this range of stellar mass, 
We calculate $M(H_2)$ of those galaxies without CO
observations from the $M(H_2)$ vs. $M_{star}$ relation as described in Boselli et al. (2013c).
The volume covered by the HRS is necessary for the normalisation. It is
calculated considering that, according to the selection criteria adopted to define the sample (Boselli et al. 2010a), we selected
galaxies in the volume between 15 and 25 Mpc over an area of 3649 sq.deg. This leads to 4539 Mpc$^3$. Given the tight relation
between the stellar mass and the molecular gas mass of galaxies within the sample (see Boselli et al. 2013c), we estimate that the HRS sample
is fairly complete for $M(H_2)$ $\gtrsim$ 5 10$^8$ M$_{\odot}$.
\\

The molecular gas distribution of the HRS galaxies can be compared to the only molecular gas distribution available in the literature.
This has been determined by Obreschkow \&
Rawlings (2009) by transforming the observed CO luminosity distribution determined from the FCRAO CO survey (Young et al. 1995) by Keres et al. (2003)
\footnote{In these works, the molecular gas mass distribution is erroneously defined molecular gas function because
the sample galaxies are not selected from a blind CO survey, but rather in the far infrared and then observed in the CO line.}.
This CO luminosity distribution has been determined for a complete sample of far-infrared selected 
late-type galaxies with a 60 $\mu$m flux density (from IRAS) $S(60\mu m)$ $\geq$ 5.24 Jy.
Obreschkow \& Rawlings (2009) converted the CO luminosity distribution of Keres et al. (2003) into an $M(H_2)$ mass distribution assuming either a constant $X_{CO}$ 
conversion factor of 2.3 10$^{20}$ cm$^{-2}$/(K km s$^{-1}$) (left panel) or a CO luminosity-dependent $X_{CO}$ 
(right panel). For this exercise, we take the molecular gas distributions derived by 
Obreschkow \& Rawlings (2009)\footnote{In the original work of Obreschkow \& Rawlings (2009) 
the conversion factor is $X_{CO}$ = 3 10$^{20}$ cm$^{-2}$/(K km s$^{-1}$).}. \\

Figure \ref{H2MF}a shows that  the molecular
gas mass distribution of the HRS galaxies determined using a constant $X_{CO}$ matches the one of the FCRAO sample only
whenever $M(H_2)$ $\gtrsim$ 6 10$^9$ M$_{\odot}$,
while it drops significantly at lower masses. Given that both ours and the FCRAO samples are complete at a given limit, we guess that
the observed difference at low $M(H_2)$ in the distribution of the two samples is due to selection effects. The CO emission of galaxies
is tightly correlated to their far-infrared emission, so we expect that the FCRAO includes galaxies with a
higher CO emission per unit stellar mass than the HRS. The K-band emission, in contrast, is due to the evolved stellar 
populations that are not necessarily associated to recent event of star formation. Quiescent galaxies, where star formation has
already transformed most of the available gas into stars, have naturally a lower molecular gas content than do star forming
systems (see Sect. 5).\\

The molecular gas mass distribution of the HRS galaxies is also very different from that of the FCRAO when molecular gas
masses are determined using a luminosity-dependent conversion factor (Fig. \ref{H2MF}b). Beyond the reasons already mentioned for the selection
criteria of the FCRAO and the HRS samples, part of the difference in the two luminosity distributions might come from the
$X_{CO}$ conversion factor, here being calibrated on the H-band luminosity, in Obreschkow \& Rawlings (2009) on the CO luminosity.
Given the wide range in CO luminosity for the FCRAO survey, the conversion factor of Obreschkow \& Rawlings (2009) used to estimate the molecular gas distribution 
in Fig. \ref{H2MF}b ranges from $X_{CO}$ = 3.7 10$^{19}$ cm$^{-2}$/(K km s$^{-1}$) to 4.1 10$^{20}$ cm$^{-2}$/(K km s$^{-1}$),
while only from $X_{CO}$ = 8.6 10$^{19}$ cm$^{-2}$/(K km s$^{-1}$) to 3.9 10$^{20}$ cm$^{-2}$/(K km s$^{-1}$) in our sample.
To conclude, Fig. \ref{H2MF} clearly indicates that the molecular gas mass distributions presented in the literature, including
the present one, can be very different by far from the real H$_2$ mass function of galaxies in the nearby universe. This is
because they strongly depend on the selection criteria adopted to define the sample and on the recipe used to convert CO
luminosities into molecular gas masses. They should thus be used with extreme caution in any comparison 
with cosmological and semi-analytic simulations.

\section{The dependence on morphological type}

\begin{figure*}
   \centering
   \includegraphics[width=18cm]{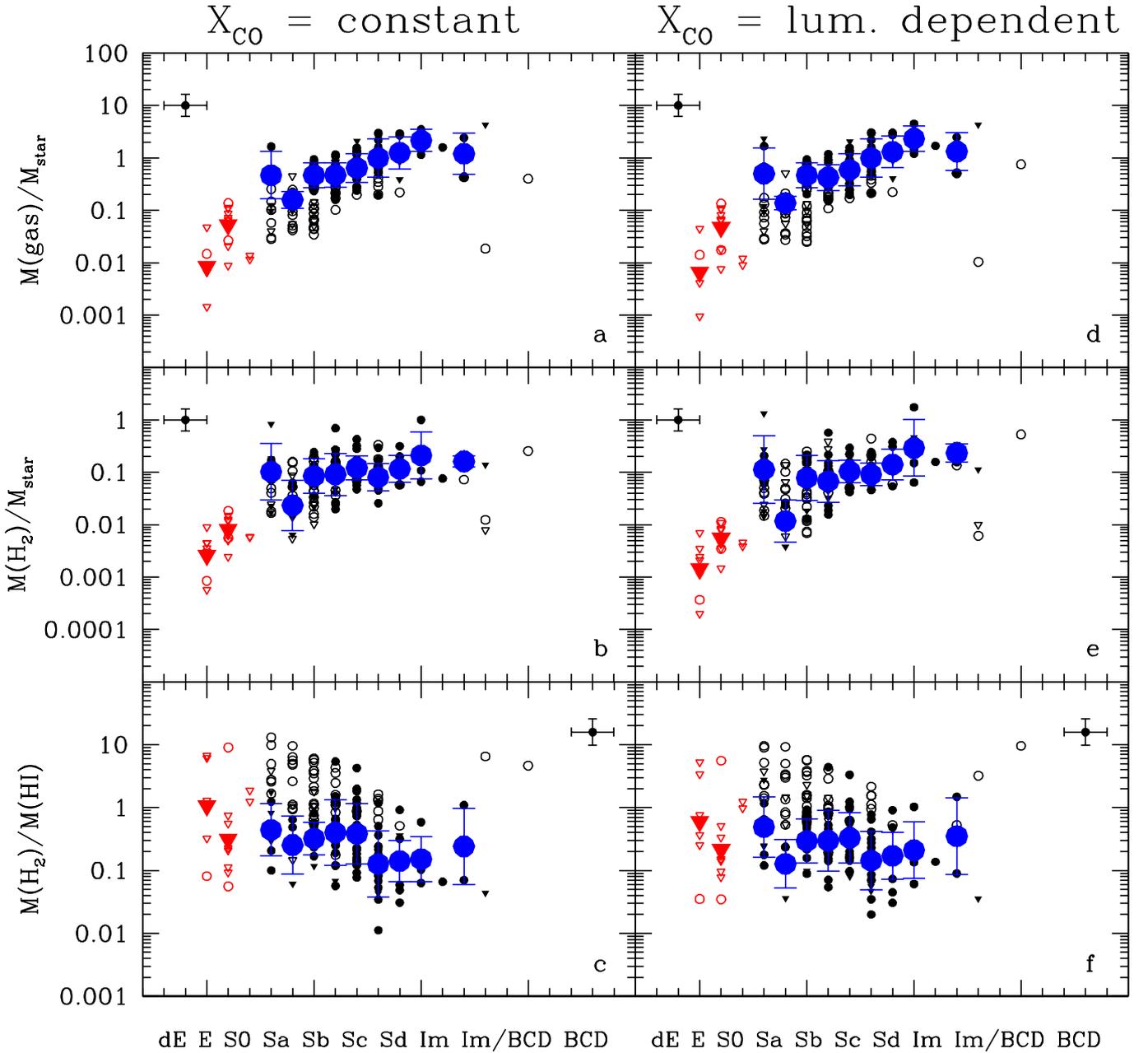}
   \caption{The relationship between the total
   gas-to-stellar mass (panels a, d), the molecular-to-stellar mass (panels b, e), and the molecular-to-atomic gas ratios (panels c, f) and morphological type. 
   The molecular gas mass is determined using a constant ($X_{CO}$ =
   2.3 10$^{20}$ cm$^{-2}$/(K km s$^{-1}$); left panels) or the H-band luminosity dependent (right panels) conversion factor of Boselli et al. (2002).
   Circles indicate detected galaxies, triangles upper
   limits. Black symbols are for late-type, red symbols for early-type HRS galaxies. For late-type systems, filled symbols indicate 
   galaxies with a normal H{\sc i} gas content ($\mathrm{\hi}-def$ $<$ 0.4), empty symbols H{\sc i}-deficient objects ($\mathrm{\hi}-def$ $\geq$ 0.4). The big symbols give the 
   mean values and the standard deviation (in log scale) in each bin of morphological type considering only H{\sc i} detected objects with $\mathrm{\hi}-def$ $\leq$ 0.4 
   and CO upper limits as detections. The typical error bar on the data is given in each panel. 
   }
   \label{gastype}%
   \end{figure*}

Figure \ref{gastype} shows the dependence of total gas (upper panel), of the molecular (middle panel) gas masses, both normalised to the stellar mass, 
and the $M(H_2)$ to $M(\mathrm{\hi})$ gas fraction (lower panel) on the morphological type. 
%Red symbols are used to identify early-type galaxies (E-S0-S0a), black symbols for late-type systems (Sa-Im-BCD).
%Triangles indicate upper limits in the molecular gas content. Late-type galaxies are coded according to their H{\sc i} gas content, proxy for the any possible perturbation induced by the cluster environment:
%filled symbols indicate unperturbed galaxies with a normal H{\sc i} gas content ($\mathrm{\hi}-def$ $<$ 0.4), empty symbols H{\sc i} gas poor objects ($\mathrm{\hi}-def$ $\geq$ 0.4). 
In the left panels the molecular gas content is measured using a constant $X_{CO}$, while in the right hand 
panels with a H-band, luminosity-dependent $X_{CO}$ conversion factor. The mean values determined considering as detection CO upper limits are given in Table \ref{Tabtype}. These values have been determined using only 
galaxies with a normal H{\sc i} content ($\mathrm{\hi}-def$ $\leq$ 0.4).
Given the large number of undetected sources, the mean values for early-type galaxies given in Table \ref{Tabtype} and plotted in Fig. \ref{gastype} must be considered as upper limits. \\

Clearly the HRS sample lacks CO data for the dwarf, metal-poor Im and BCD galaxies, where CO observations are known to be
challenging. Despite this evident weakness in the dataset, Fig. \ref{gastype} shows that the molecular gas content per unit
stellar mass is fairly constant along the spiral sequence ($M(H_2)$ $\simeq$ 0.1 $M_{star}$ for all late-type galaxies of type Sa-Im), 
while it drops drastically in early-type objects (E-S0a; $M(H_2)$ $\lesssim$ 0.01 $M_{star}$; panels b, e). 
The total cold gas phase, in contrast, increases by more than a factor of 10 from early spirals (Sa-Sb) to late dwarfs (Sd-Im-BCD): 
$M(gas)$ is $\lesssim$ 1\% ~ of the stellar mass in early-type objects, is in between $\sim$ 10 and 100\% of $M_{star}$
in late-type galaxies (Sa-Sc), and exceeds the stellar mass by a factor of 2-4 in Sm-Im objects, thus becoming here the dominant baryonic component
of galaxies (panels a, d). The total molecular gas content is $M(H_2)$ $\simeq$ 0.1-0.5 $M(\mathrm{\hi})$. The mean value of log$M(H_2)/M(\mathrm{\hi})$ in late-type, 
non-deficient ($\mathrm{\hi}-def$ $<$ 0.4) H{\sc i}-detected
galaxies (determined considering CO upper limits as detections) is log($M(H_2)/M(\mathrm{\hi})$)=-0.59$\pm$0.50 whenever $M(H_2)$ is calculated with a constant conversion factor,
log($M(H_2)/M(\mathrm{\hi})$)=-0.61 $\pm$ 0.46 when $X_{CO}$ is luminosity dependent. The same variables are log($M(H_2)/M(\mathrm{\hi})$)=-0.54$\pm$0.51 for $X_{CO}$ constant,
log($M(H_2)/M(\mathrm{\hi})$)=-0.57 $\pm$ 0.45 for $X_{CO}$ variable if CO non-detections are not considered. The molecular gas content is thus $\sim$ 25-30 \% ~ of the atomic gas, consistent with 
our previous results (Boselli et al. 2002; Saintonge et al. 2011a).

The lack of any strong dependence of the molecular gas content on morphological type for late-type systems
mainly confirms similar results obtained on different sets of data by Young \& Knezek (1989), Young \& Scoville (1991), Sage (1993), Boselli et al. (1997, 2002), 
Casoli et al. (1998), Sauty et al. (2003), and more recently Lisenfeld et al. (2011). The quality of our data,
the accurate estimate of the uncertainties on the total CO flux extrapolation, and the use of a constant or a luminosity-dependent conversion factor
on the same set of data allow us to reduce the scatter in the observed distributions.\\ 

The dependence of the total gas content and of the H$_2$ to H{\sc i} gas fraction on Hubble type is also consistent with what hasbeen
observed by Young \& Knezek (1989), Young \& Scoville (1991), Sage (1993), Boselli et al. (1997, 2002), Casoli et al. (1998), Sauty et al. (2003), and Lisenfeld et al. (2011). 
Obreschkow \& Rawlings (2009) used the FCRAO survey of Young et al. (1995) to observe a gradual decrease in the  $M(H_2)/M(\mathrm{\hi})$ ratio as a function of morphological type (from
$M(H_2)/M(\mathrm{\hi})$ $\simeq$ 0.6 in Sa to $M(H_2)/M(\mathrm{\hi})$ $\simeq$ 0.1 in Sm)
a trend not confirmed by our set of data.\\

\section{The scaling relations}

\subsection{Gas to stellar mass scaling relations}

\begin{figure*}
   \centering
   \includegraphics[width=18cm]{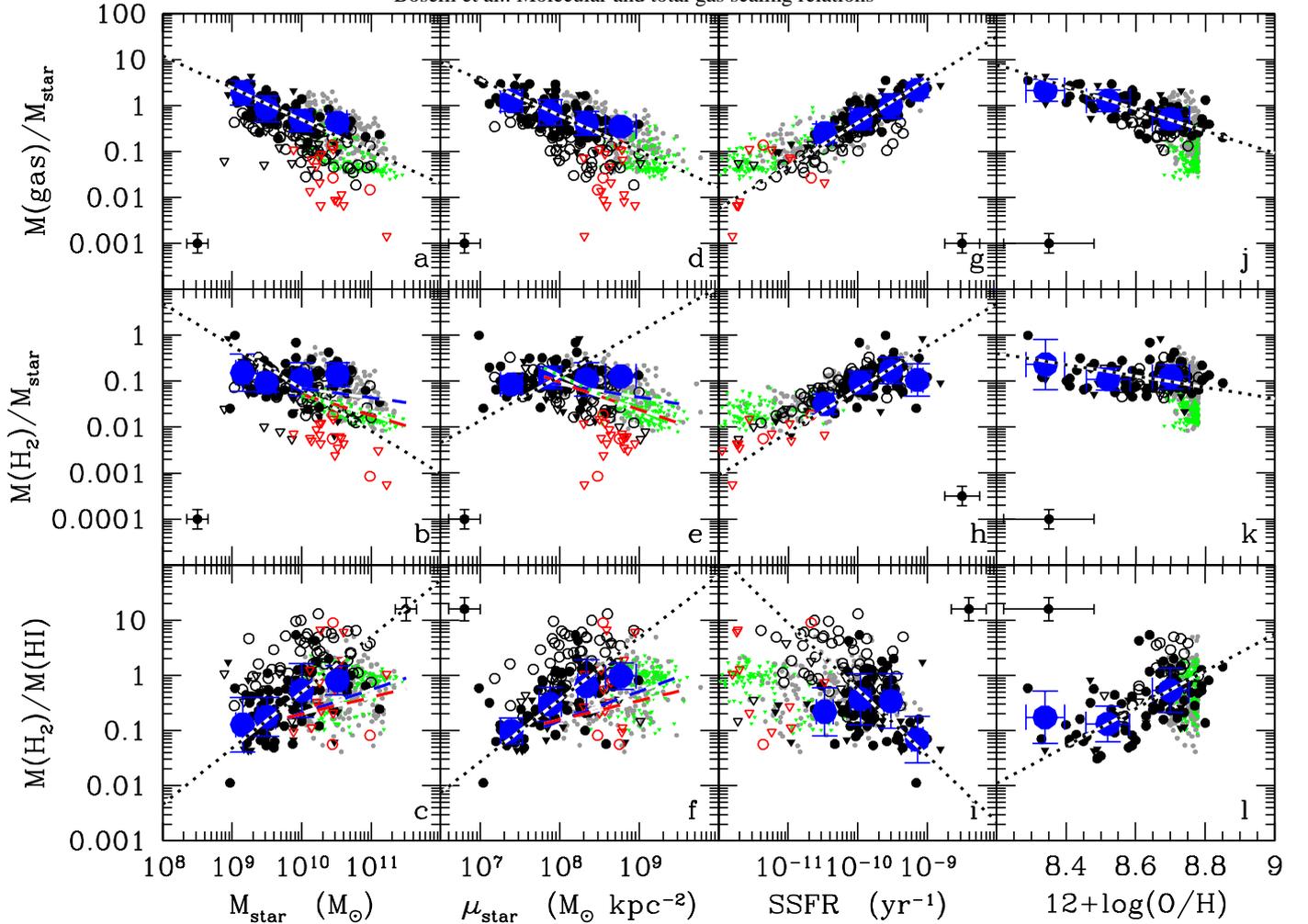}
   \caption{The scaling relation concerning the total gas-to-stellar mass ratio (upper rows), the molecular gas-to-stellar mass ratio (middle rows) and
   the molecular-to-atomic gas ratio (lower rows). These entities are plotted vs. the total stellar mass (left panels), the stellar surface density (middle-left panels),
   the specific star formation rate (middle-right panels), and the metallicity (right panels). Molecular gas masses are determined assuming a constant ($X_{CO}$ =
   2.3 10$^{20}$ cm$^{-2}$/(K km s$^{-1}$)) conversion factor.
   Circles indicate detected galaxies, triangles upper
   limits. Black symbols are for late-type, red symbols for early-type HRS galaxies, grey symbols for detected and green symbols for undetected (upper 
   limits) COLD GASS objects (these last corrected for the same conversion factor). For the HRS, filled symbols indicate late-type
   galaxies with a normal H{\sc i} gas content ($\mathrm{\hi}-def$ $<$ 0.4), empty symbols H{\sc i}-deficient objects ($\mathrm{\hi}-def$ $\geq$ 0.4). The big blue symbols give the 
   mean values and the standard deviation (in log scale) considering only H{\sc i} and CO detected late-type objects.  
   The typical error bars on the data are given in each panel. 
   These are errors on the total CO flux and do not include any other error in the transformation of the CO fluxes into molecular gas
   masses.  
   The black dotted line indicates the bisector fit determined using only
   detected, late-type gas-rich galaxies ($\mathrm{\hi}-def$ $<$ 0.4; black filled dots). The blue dashed, red dashed, and green dotted lines show the best fit 
   obtained by Saintonge et al. (2011a) for the COLD GASS sample when considering only CO detections (blue), CO upper limits as detections (red), and CO
   undetections as zero values (green).  }
   \label{scalingcos}%
   \end{figure*}

\begin{figure*}
   \centering
   \includegraphics[width=18cm]{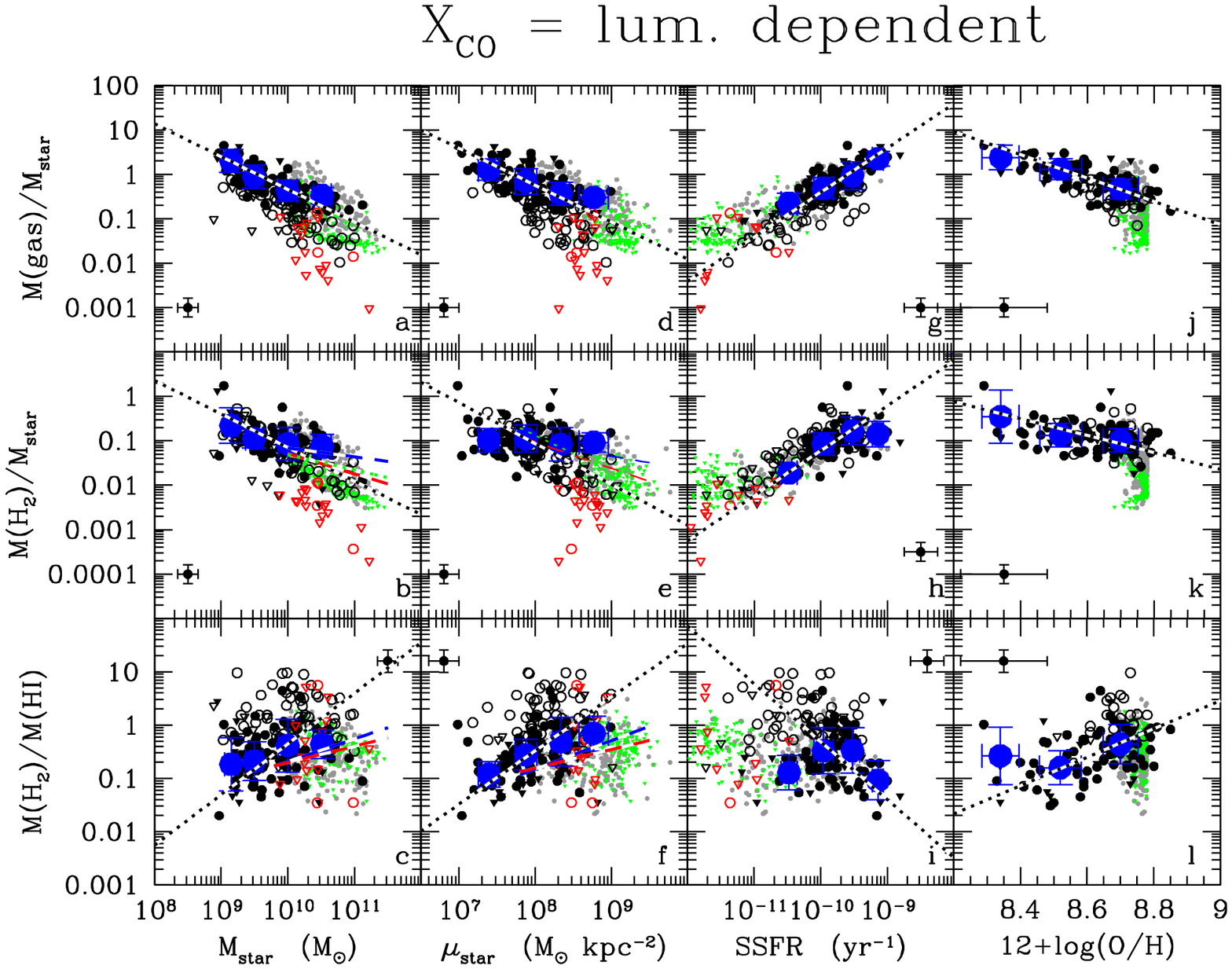}
   \caption{Same as Fig. \ref{scalingcos} when molecular gas masses are estimated using the H-band luminosity-dependent conversion factor of Boselli et al. (2002). }
   \label{scalingvar}%
   \end{figure*}

Figures \ref{scalingcos} and \ref{scalingvar} show the scaling relations relative to the total gas-to-stellar mass ratio (upper panels), 
the molecular gas-to-stellar mass ratio (middle panels),
and the molecular-to-atomic gas ratio (lower panels). These quantities 
are plotted vs. the total stellar mass, the stellar mass surface density,
the specific SFR and the metallicity. In Fig. \ref{scalingcos} the molecular gas content is estimated assuming a constant conversion factor, while in Fig. 
\ref{scalingvar} assuming the H-band luminosity dependent $X_{CO}$ of Boselli et al. (2002).
The best fit to the data done using only unperturbed HRS field late-type galaxies ($\mathrm{\hi}-def$ $<$ 0.4) and excluding all COLD GASS objects 
are given in Table \ref{Tabscalingfit} (bisector fit), while the mean values 
in different bins for the same subsample of objects are listed in Table \ref{Tabscalingdata}. Only H{\sc i} and CO detections are considered here.
In this section we discuss the scaling relations relative to the total gas and molecular gas to stellar mass ratios, while those concerning $M(H_2)/M(\mathrm{\hi})$ are
presented in the next section. \\
Overall Figs. \ref{scalingcos} and \ref{scalingvar} show fairly defined trends between $M(gas)/M_{star}$ 
and $M_{star}$, $\mu_{star}$, $SSFR$, and 12+log(O/H), as well as between $M(H_2)/M_{star}$ and $SSFR$. The total gas-to-stellar mass ratio decreases with increasing stellar mass, stellar mass surface density, and metallicity
and increases with increasing specific SFR. $M(H_2)/M_{star}$ also increases with the specific SFR. 
A detailed analysis of the same figures and of Tables \ref{Tabscalingfit} and \ref{Tabscalingdata}, however,
indicates the following:\\
1) The relationships concerning $M(H_2)/M_{star}$ are always flatter and more scattered than those with $M(gas)/M_{star}$. Table \ref{Tabscalingfit} 
clearly indicates that all the scaling relations determined using the total gas-to-stellar mass ratio are statistically significant (Spearman correlation coefficient $\rho$ $\geq$ 0.64).
Among those concerning $M(H_2)/M_{star}$, the only one statistically significant is with the specific SFR ($\rho$ $\geq$ 0.47). 
The larger scatter in the molecular gas scaling relations with respect to those with the total gas partly comes from the large uncertainty 
in the determination of the total molecular gas mass ($\sim$ 44 \%), 
which is by far much more uncertain than that of the atomic gas phase ($\sim$ 15 \%). \\
2) %We try to quantify any systematic effect related to 
%the uncertainty on the conversion factor assuming either a constant (Fig. \ref{scalingcos}) or a luminosity dependent (Fig \ref{scalingvar}) 
%$X_{CO}$ in the determination of the molecular gas mass.
The scaling relations where gas masses are determined using a variable $X_{CO}$ are always steeper and more significant than those when $M(H_2)$ is determined with
a constant $X_{CO}$. This effect is obviously less pronounced whenever total gas masses are used. 
%The anti-correlation between $M(H_2)/M_{star}$ vs. $M_{star}$, evident when 
%$M(H_2)$ is determined using a luminosity dependent $X_{CO}$ ($\rho$ = -0.60), for instance is almost removed when $X_{CO}$ is taken constant ($\rho$ = -0.16). 
Among the molecular hydrogen scaling relations, the correlation of $M(H_2)/M_{star}$ on $SSFR$ is the only valid one regardless of the use of a constant or a variable conversion factor.
All scaling relations concerning the total (molecular plus atomic) gas content are statistically significant and show little variation with $X_{CO}$.\\
3) The dynamic range covered by these relations extends the one sampled by the COLD GASS sample to lower stellar masses (from 10$^{10}$ $\lesssim$ $M_{star}$ $\lesssim$ 2 10$^{11}$ M$_{\odot}$
to 10$^{9}$ $\lesssim$ $M_{star}$ $\lesssim$ 10$^{11}$ M$_{\odot}$) and lower stellar mass surface densities (from 10$^{8}$ $\lesssim$ $\mu_{star}$ $\lesssim$ 3 10$^{9}$ M$_{\odot}$ kpc$^{-2}$
to 10$^{7}$ $\lesssim$ $\mu_{star}$ $\lesssim$ 10$^{9}$ M$_{\odot}$ kpc$^{-2}$). Unperturbed, late-type HRS galaxies have specific SFR in the range
2 10$^{-11}$ $\lesssim$ $SSFR$ $\lesssim$ 10$^{-9}$ yr$^{-1}$, while perturbed cluster objects extend down to $SSFR$ $\simeq$ 10$^{-11}$ yr$^{-1}$. The HRS galaxies are thus generally 
more active than a large fraction of the quiescent, massive early-type COLD GASS objects, where the specific SFR extends down to
$SSFR$ $\simeq$ 10$^{-12}$ yr$^{-1}$. The combination of HRS and COLD GASS data 
is thus crucial for determining the validity of the observed scaling relations on a wide and significant range in the parameter space. 
In the overlap region, the HRS and COLD GASS samples are fairly consistent. We observe, however, a discontinuity in the $M(H_2)/M_{star}$ vs.
$M_{star}$ and $\mu_{star}$ scaling relations when combining the HRS sample to COLD GASS detected galaxies. The very few massive galaxies of the HRS sample seem
consistent with the COLD GASS data. The slope of the fitted relations thus changes 
from the HRS to the COLD GASS sample in the $M(H_2)/M_{star}$ vs. $M_{star}$ and $\mu_{star}$ relations\footnote{Here we calculate bisector
fits, while those given in Saintonge et al. (2011) are linear fits}.
The difference in the fitted relations can thus be due to the narrow dynamic range of the COLD GASS
sample. We do not confirm the anti-correlation between $M(H_2)/M_{star}$ vs. $M_{star}$ claimed by Saintonge et al. (2011a). Indeed this relation is present
within the HRS only when $M(H_2)$ is determined using a luminosity-dependent $X_{CO}$ ($\rho$ = -0.52), while it is not statistically significant when $X_{CO}$ 
is taken to be constant ($\rho$ = -0.03). \\
4) The very stringent upper limits determined for the HRS clearly indicate that early-type galaxies (E-S0-S0a)
do not follow the same scaling relations as in late-type systems\footnote{The upper limits in the molecular and atomic gas of the undetected HRS galaxies are determined
consistently with those of the COLD GASS survey.}.\\ 

Since these scaling relations are driven by the formation and evolution process that gave birth to local, evolved late-type galaxies, 
they can be used to constrain their star formation history. The relationship between the total gas-to-stellar mass ratio vs. metallicity, for instance, 
is a well known relation expected 
for the chemical evolution of galaxies (Searle \& Sargent 1972, Edmunds 1990, Garnett 2002, Dalcanton 2007).
The observational evidence collected so far indicates that, although locally the 
process of star formation is mainly related to the molecular gas (Schruba et al.
2011), the total gas content is the gaseous phase that mainly drives galaxy evolution. This result is quite striking since a large portion of the atomic gas, which
dominates the gaseous phase of the ISM, is located in the outer disc of galaxies. 
This gas is not necessarily associated to star forming regions since located outside the optical disc. 

The contribution of the UV extendend discs to the total star formation activity of galaxies is indeed minor (Thilker et al. 2007).
In normal, unperturbed galaxies, the H{\sc i} gas has a fairly flat distribution which extends up to $\sim$ 1.8 optical radii (Cayatte et al. 1994).
It seems thus that, even though locally the star formation process is mainly related to the molecular gas component (Bigiel et al. 2008), the global star formation
history of late-type galaxies is instead driven by the total gas content, which includes a large portion of gas not directly participating to the star
formation process (Sancisi et al. 2008; Fraternali \& Tomassetti 2012). This evidence, already noted in Boselli et al. (2001), suggests that there is a sort of physical relation between the gas present in giant molecular
clouds and star forming regions and the diffuse component located in the outer disc. It might be indirect evidence of the feedback process of star formation in 
galactic discs. As discussed in Boselli et al. (2001), this evidence is consistent with a secular evolution of galaxies, 
where the star formation activity has been rapid at early epochs in massive objects while it is still present at a rate comparable 
to the mean rate since their formation in low mass systems (Gavazzi et al. 1996; Cowie et al. 1996; 
Boissier \& Prantzos 1999, 2000).

The second interesting result is that all the scaling relations analysed in this work relative to the $M(H_2)/M_{star}$ ratio are steeper whenever $M(H_2)$ is determined using 
a luminosity dependent than a constant CO-to-H$_2$ conversion factor. This result is expected given the tight relation between stellar mass, stellar mass surface density,
colour, specific SFR, metallicity, and the H-band luminosity (Gavazzi et al. 1996; Boselli et al. 2001), the parameter used to derive the luminosity-dependent 
$X_{CO}$ for each target galaxy. The debate over any possible systematic variation
in the CO-to-H$_2$ conversion factor on the physical properties of the ISM, such as the metallicity or the intensity of the interstellar radiation field, 
is still far from being settled both observationally and theoretically (Boselli et al. 2002, Bell et al. 2007, Bolatto et al. 2008, Liszt et al. 2010, Leroy et al. 2011, 
Shetty et al. 2011a,b, Narayanan et al. 2012, Kennicutt \& Evans 2012, Bolatto et al. 2013, Sandstrom et al. 2013). We thus cannot state from the present 
analysis which is the most reliable between the two sets of relations. We are inclined to consider the 
difference between the relations determined using a constant and a luminosity-dependent conversion factor as a
realistic estimate of the uncertainty on the relationships themselves. This can, of course, be extended to all the relations considered in the present work
using $M(H_2)$ as a variable ($M(H_2)/M(\mathrm{\hi})$, gas depletion timescales).\\ 

\subsection{Molecular-to-atomic gas scaling relations}

The molecular-to-atomic gas ratio is sensitive to the physical process of transformation of the two gas phases in the ISM.
It is thus an interesting parameter for constraining the physical properties of the interstellar medium. Figures \ref{scalingcos} and  
\ref{scalingvar} show the $M(H_2)/M(\mathrm{\hi})$ scaling relations for the HRS and compare them to those determined for the COLD GASS sample.
We plot here only H{\sc i}-detected galaxies with upper limits on $M(H_2)/M(\mathrm{\hi})$.
The best fit to the relations (bisector fit) are given in 
Table \ref{Tabscalingfit}, with the mean values in different bins in Table \ref{Tabscalingdata}. Fits and mean values are determined using the subsample of
unperturbed ($\mathrm{\hi}-def$ $<$ 0.4) late-type HRS galaxies detected in H{\sc i} and CO. These figures indicate that\\
1) $M(H_2)$/$M(\mathrm{\hi})$ increases with $M_{star}$, $\mu_{star}$, 
and 12+log(O/H), while it decreases with $SSFR$. They thus often show the opposite trend to those observed for the $M(H_2)$/$M_{star}$ and $M(gas)/M_{star}$ ratios. \\
2) The relations are steeper when a constant $X_{CO}$ conversion factor is used. Among these, the only relations that are statistically significant regardless of the use of
a constant or a variable conversion factor are those with stellar mass surface density and metallicity.\\
3) Compared to the relations determined using the COLD GASS sample, those determined for the HRS are 
slightly steeper. The difference in slope is more pronunced when, consistently with Saintonge et al. (2011a), the fit is done 
on molecular gas masses determined assuming a constant $X_{CO}$\footnote{We recall that in COLD GASS the fits are linear fits, while bisector fits here.}. As for the previously discussed relations, this difference
probably comes from the narrower range in the parameter space covered by the COLD GASS sample with respect to the HRS. The COLD GASS galaxies not sampled by the HRS (the
most massive objects with high stellar surface densities and low SSFR) have, on average, $M(H_2)$/$M(\mathrm{\hi})$ values
lower than those extrapolated from the scaling relations determined from the HRS. \\
4) The few early-type galaxies detected in H{\sc i} do not follow the same scaling relations than late-type systems.\\
5) H{\sc i} deficient galaxies have high $M(H_2)$/$M(\mathrm{\hi})$ ratios and do not follow the scaling relations traced by normal, unperturbed galaxies. 

The physical interpretation of the dependence of $M(H_2)/M(\mathrm{\hi})$ on 12+log(O/H) is not straightforward. Locally we expect a tight
connection between the molecular and the atomic gas phase given that the molecular gas is generally formed through the condensation of atomic hydrogen on dust grains
(Hollenbach \& Salpeter 1971). Recent models of Krumholz et al. (2009), indeed, predict a dependence of the molecular to atomic hydrogen mass ratio on  
the dust column density, thus indirectly on the metallicity or the total gas column density of the parent galaxy. 
These relations are theoretically justified by the fact that there is an equilibrium between the amount of molecular gas formed on dust grains 
and that dissociated by the interstellar radiation field in photodissociation regions on the outskirts of molecular clouds (Elmegreen 1993). The total gas column density 
is a key parameter because the molecular gas can shield itself from the interstellar radiation field, acting thus as a screen preventing the molecular gas from
photodissociation (Wolfire et al. 1995). The gas metallicity is a direct indicator of dust attenuation and dust content within molecular clouds. Again, whenever the attenuation is high,
which is generally the case when the molecular phase is dominant (Boquien et al. 2013), the interstellar radiation field cannot penetrate and 
dissociate the molecular gas phase and thus reduce the $M(H_2)/M(\mathrm{\hi})$ ratio.\\

\subsection{Comparison with models of galaxy formation and evolution}

The most recent simulations of galaxy formation include the atomic and molecular gas phases in the
interstellar medium of galaxies
(e.g. Fu et al. 2010; Lagos et al. 2011a; Duffy et al. 2012).
The scaling relations presented in the
previous sections, combined with those given in Cortese et al. (2011)
for the atomic gas component,
represent strong constraints for these models.
We qualitatively discuss how model predictions compare to the observations, with the aim
of determining
whether the
physical prescriptions included in the models are consistent with the observations of
local galaxies or not.\\

Two broad prescriptions to account for the transformation from the atomic to molecular
gas have been included in galaxy formation semi-analytic models and simulations.
The first one is based on the models of
Krumholz et al. (2009) where the formation of molecular gas mainly depends
on the strength of the interstellar radiation
field and on the dust opacity of clouds. The second prescription assumes that the atomic
to molecular gas transformation is regulated by the midplane disc hydrostatic
pressure.
These prescriptions have observational and theoretical support.
The simulations of Gnedin et al. (2009) indicate that the molecular phase depends
primarily on metallicity, as is indeed observed in the HRS galaxies. This arises from
an underlying correlation between temperature and the chemical state of the ISM (Schaye 2004).
Such correlations are expected to break down at metallicities lower than
$10^{-2}\,Z_{\odot}$ (e.g. Glover et al. 2012), which is well below the metallicities
of the HRS galaxies.\\

Fu et a. (2010) and Lagos et al. (2011b) include the two broad prescriptions
described above in semi-analytic models of galaxy formation and
predict that the atomic gas
fraction $M(\mathrm{\hi})/M_{star}$ is inversely correlated with stellar mass, stellar mass surface
density, and gas mass surface density.
They also predict that the molecular gas fraction, $M(H_2)/M_{star}$, is correlated with the
gas mass surface density but shows only a mild dependence on stellar mass or
stellar mass surface density. The $M(H_2)/M(\mathrm{\hi})$ ratio should then be tightly correlated with
gas surface density and weakly with stellar mass and stellar mass surface density.
Lagos et al. (2011b) also show that there is a weak trend between morphological type and
the $M(H_2)/M(\mathrm{\hi})$ ratio, where early-type galaxies have greater $M(H_2)/M(\mathrm{\hi})$ ratios than
late-type galaxies, due to the contribution from the stellar surface density to the midplane
disc pressure in the former.
These relations are predicted for galaxies
with stellar masses in the range 10$^{9}$ $\lesssim$ $M_{star}$ $\lesssim$ 10$^{12}$ M$_{\odot}$
in the case of Fu et al., and 10$^{8}$ $\lesssim$ $M_{star}$ $\lesssim$ 10$^{12}$ M$_{\odot}$
in the case of Lagos et al. \\

The predictions of Fu et al. (2010) and Lagos et al. (2011b) are fairly consistent
with the general trends observed in the molecular gas fraction scaling
relations in the HRS galaxies. They are also consistent with those relative to the
atomic gas fraction presented in Cortese et al.
(2011) for the same sample of galaxies. They also reproduce the observed
trends fairly well between the molecular-to-atomic gas phase and
the total stellar mass and the total stellar mass surface density.
However, recent detailed comparisons between the predictions of the model of
Fu et al. (2010) and the COLD GASS sample, presented by Kauffmann et al. (2012), show
that the models fail to reproduce the
very small dispersion of some of the observed scaling relations. Kauffmann et al. interpret
this discrepancy as arising from the feedback mechanisms included in the models, which efficiently
reduce the gas content and quench star formation
in galaxies with stellar masses $M_{star}$ $\gtrsim$ 10$^{10}$ M$_{\odot}$. In the observations,
such a sharp decrease in the gas fractions of massive galaxies is not present, and instead
the stellar surface density seems to be more indicative of quenching.

Similar conclusions have been reached by Duffy et al. (2012) in hydrodynamical simulations of galaxy formation.
Duffy et al. include the pressure prescription to calculate the atomic and molecular
hydrogen content of galaxies and show that
their predicted scaling relations agree with the observations only when AGN feedback
is not included in the simulation. On the other hand, stellar feedback plays a key role in lowering the gas
fractions and bringing the models back into agreement with the observations (see also Dave et al. 2013).
This conflicts with the stellar mass function, in which the massive-end is largely overestimated if AGN
feedback is not included.\\

From the qualitative comparison presented in this section,
it is fair to say that the prescriptions used in semi-analytic models and simulations
describing the content of atomic
and molecular gas in galaxies offer a feasible physical framework
to explain the observed gas scaling relations due to the broad agreement between the model predictions and
the observations. However, the models still need to explore and
improve their treatment of feedback mechanisms in order to help understand
the gas fraction distributions of galaxies with different stellar masses and
morphologies.\\

\section{The gas depletion timescale}

\begin{figure*}
   \centering
   \includegraphics[width=18cm]{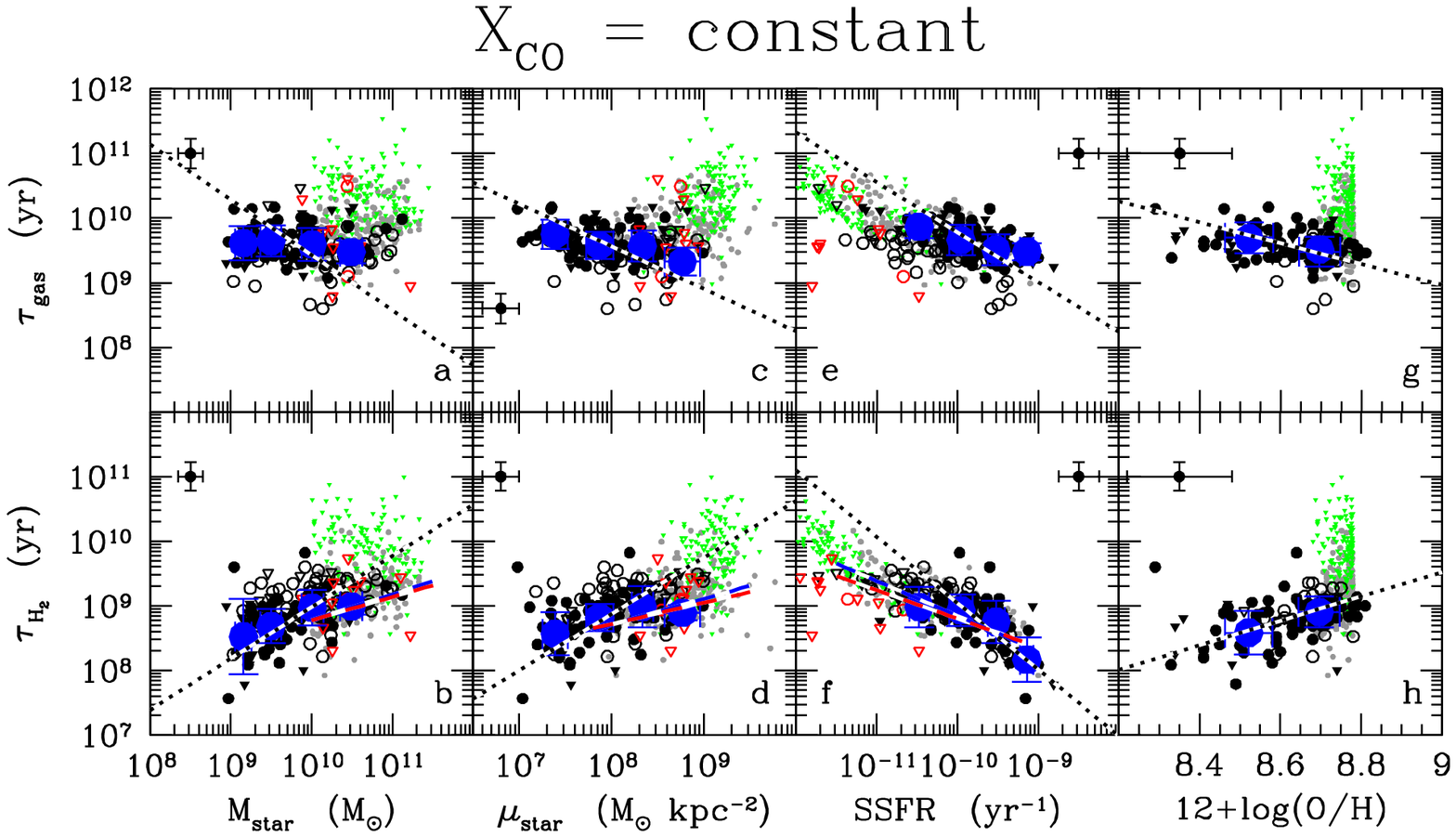}
   \caption{The relationship between the total (upper row) and the molecular (lower row) gas depletion timescale and stellar mass 
   (left panels), the stellar mass surface density (middle-left panels), the specific star formation rate (middle-right panels), and the metallicity 
   (right panels). Molecular gas masses are determined assuming a constant ($X_{CO}$ =
   2.3 10$^{20}$ cm$^{-2}$/(K km s$^{-1}$)) conversion factor.
   Same symbols as Fig. 5}
   \label{taucos}%
   \end{figure*}

\begin{figure*}
   \centering
   \includegraphics[width=18cm]{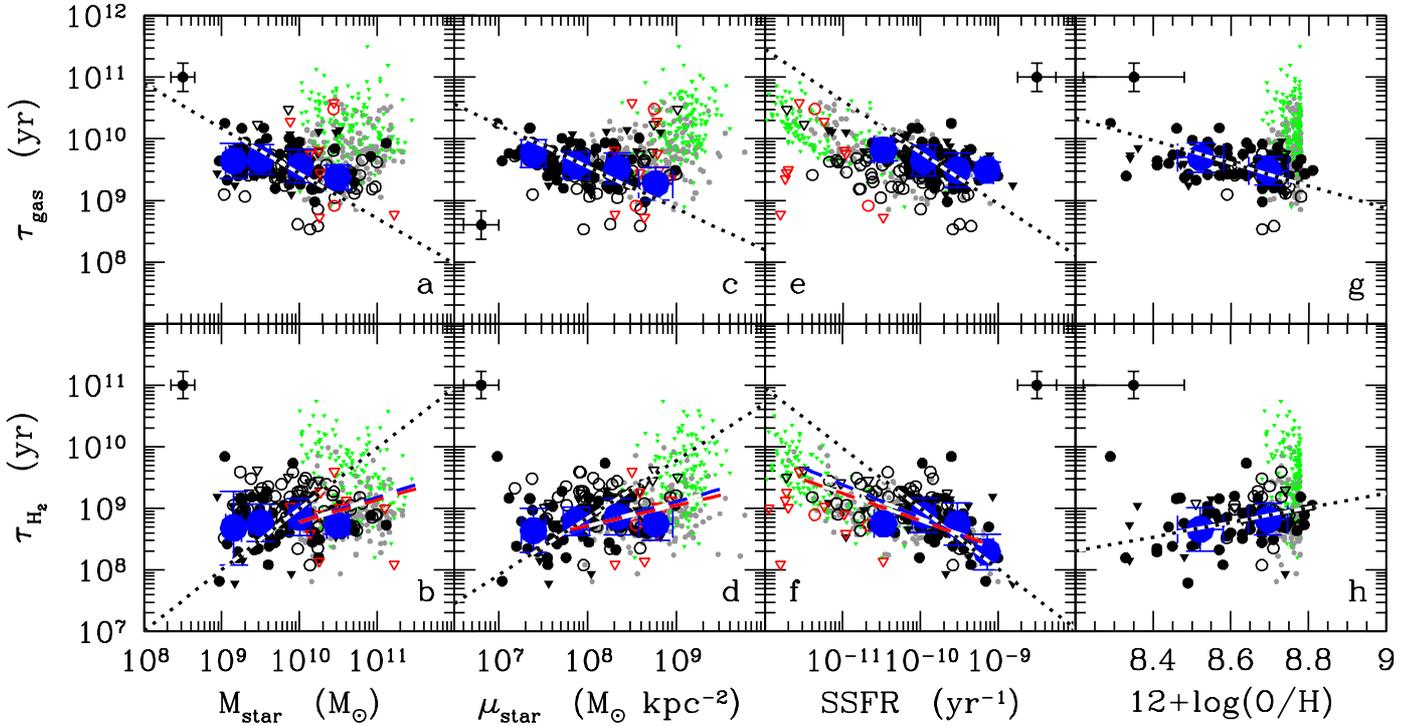}
   \caption{Same as Fig. \ref{taucos} when molecular gas masses are estimated using the H-band luminosity-dependent conversion factor of Boselli et al. (2002). }
   \label{tauvar}%
   \end{figure*}

Figures \ref{taucos} and \ref{tauvar} show the relationships between the total (upper panels) and the molecular (lower panels)
gas depletion timescales, $\tau_{gas}$, $\tau_{H_2}$, and $M_{star}$, $\mu_{star}$, $SSFR$, and 12+log(O/H) 
when molecular gas masses are determined assuming a constant or an H-band luminosity-dependent 
conversion factor. The best fit to the relations, as well as the mean values in given bins of the X-axis variables, are listed in 
Tables \ref{Tabtaufit} and \ref{Tabtaudata}, respectively. Only H{\sc i} and CO detected galaxies are considered. The analysis of Figs. \ref{taucos} and \ref{tauvar} and
Tables \ref{Tabtaufit} and \ref{Tabtaudata} indicates that\\
1) The total gas depletion timescale seems to decrease with increasing stellar mass, stellar mass surface density, SSFR, and metallicity.
The molecular gas depletion timescale does not always follow the same trends.
Among the plotted relations, however, the only variables that are correlated on a statistical basis are $\tau_{H_2}$ and $SSFR$ ($\rho$ $\simeq$ -0.55) and, to a 
lower extent, $\tau_{gas}$ and $SSFR$ ($\rho$ $\simeq$ -0.45), and $\tau_{gas}$ and 12+log(O/H) ($\rho$ $\simeq$ -0.35). \\
2) Figure \ref{tauvar} shows that the trend between $\tau_{H_2}$ vs. $M_{star}$ ($\rho$=0.52) claimed by Saintonge et al. (2011b) on the COLD GASS sample,
indicating that the molecular gas depletion timescale increases with stellar mass, is also present in the HRS sample and extends to stellar masses
as low as 10$^9$ M$_{\odot}$ when the molecular gas is determined assuming a constant conversion factor. This trend, however, is totally 
removed (if not inversed) whenever $X_{CO}$ is luminosity dependent, indicating that if a trend between the two variables exists, it is very marginal.
Similarily, the trends between $\tau_{H_2}$, $\mu_{star}$ and 12+log(O/H) observed when the molecular gas is measured using a constant $X_{CO}$ factor are 
totally removed whenever $X_{CO}$ is luminosity dependent. \\

The star formation efficiency, or equivalently the gas depletion timescale as defined in Sect. 3, gives a quantitative estimate of the efficiency
with which the gas reservoir, which is the principal feeder of star formation, is transformed into stars. Similarily, the gas depletion timescale gives an estimate
of the time that the galaxy has to sustain star formation at a rate similar to the present one when evolving as a closed box. The mean values of gas depletion timescale for 
the gas-rich ($\mathrm{\hi}-def$ $\leq$ 0.4) HRS galaxies is log($\tau_{H_2}$) $\simeq$ 8.80$\pm$0.35 yr 
and log($\tau_{gas}$) $\simeq$ 9.60 $\pm$ 0.25 yr, which correspond to $\tau_{H_2}$ $\simeq$ 0.6 Gyr and $\tau_{gas}$ $\simeq$ 4.0 Gyr. 
These values can be compared to the mean value determined from the HERACLES sample of resolved galaxies by Leroy et al. (2008, 2013), $\tau_{H_2}$ $\simeq$ 1.3-2.2 Gyr,
measured on a typical 800 pc scale, or to those determined for different samples of galaxies as listed in Table 5 of Leroy et al. (2013). As underlined in Sect. 3, however, these are just crude
estimates because these measures do not consider that a large portion of the stellar mass is re-injected into the interstellar medium through stellar
winds (recycled gas $R$ $\simeq$ 30\%; Kennicutt et al. 1994). Furthermore we expect that the star formation of normal, late-type galaxies, such as those 
composing the HRS sample, is not constant over cosmic time but smoothly changes following different paths according to the morphological
type (Sandage 1986) or the total mass of the parent galaxy (e.g. Gavazzi et al. 1996, 2002; Boselli et al. 2001). These smooth variations with time of
the star formation activity are also predicted by chemo-spectrophotometric models of galaxies undergoing a secular evolution (Boissier \& Prantzos 1999;
2000). Smooth variations are indeed expected because of the tight relation between the star formation
and the gas surface density known as the Schmidt law (Schmidt 1959; Kennicutt 1998). The amount of gas available to sustain star formation thus depends
on different factors: on the total gas reservoir, whose content is related to the evolution of the dark matter halo in which the galaxy resides; on the past 
star formation history, which is regulated during time the transformation of gas into stars and indirectly the re-injection of the recycled gas into the ISM; on the
infall of fresh gas on the galactic disc (Tinsley 1980, Pagel 1997); and on the outflow induced by AGN in massive galaxies (Cattaneo et al. 2009)
and supernovae in dwarf systems (Larson 1976, Tenorio-Tagle \& Bodenheimer 1988). For these reasons the gas depletion timescales reported above only gives a very crude estimate of 
the time during which galaxies can continue their activity of star formation.\\ 

We do not confirm the dependence of $\tau_{H_2}$ on the stellar mass of galaxies observed by Saintonge et al. (2011b) just because the weak trend
observed when the molecular gas mass is calculated using a constant conversion factor is totally removed (or even inversed) when a luminosity-dependent
$X_{CO}$ is adopted. Consistently with Leroy et al. (2013), we do not see any statistically significant relation between the $\tau_{H_2}$ and the stellar 
surface density. The weak trend observed between $\tau_{H_2}$ and 12+log(O/H) is also consistent with what is reported by Leroy et al. (2013) on resolved galaxies. 
The trend observed between $\tau_{H_2}$ and $SSFR$, in contrast, is statistically significant ($\rho$ $\simeq$ 0.6) even though 
the relations are quite weak (but steeper than those observed by Saintonge et al. (2011b) for massive galaxies). This trend is also consistent with the one observed by Leroy et al. (2013)
on resolved galaxies. It must be noted that
$\tau_{H_2}$ and $SSFR$ are not fully independent variables since they both depend on the SFR, which is present in the denominator in the former and
in the numerator in the latter variable.
The anti-correlation between the total gas consumption timescale and $M_{star}$, $\mu_{star}$, $SSFR$, and 12+log(O/H), 
if present, are also very weak. They depend much less on
any assumption on the adopted $X_{CO}$ conversion factor just because the molecular phase is a small fraction of the total gas reservoir. 
It is interesting to note that in some cases the observed relations between $\tau_{gas}$ and the other variables are opposite to those 
observed when the molecular gas depletion timescale is considered.

%\subsection{Relation with the Schmidt law}

As previously mentioned, the dependence of $\tau$ (or equivalently $SFE$) on the different physical parameters used in this analysis are tightly connected to the
Schmidt law (Kennicutt 1998):

\begin{equation}
{\Sigma(SFR) = A \Sigma^N(gas)}
\end{equation}

\noindent
where $A$ gives the efficiency with which the gas is locally transformed into new stars. For an exponent $N$=1, 
$A$ roughly corresponds to the $SFE$ or, equivalently to 1/$\tau$\footnote{The star formation efficiency $SFE$ =1/$\tau$ deduced from eq. (5) and (6) is determined using total
gas masses and SFRs, while $A$ in eq. (12) using gas and SFR surface densities.}.
In his seminal work, Kennicutt (1989, 1998) has shown using integrated data that $N$ = 1.4 $\pm$0.15 when the total gas surface density is assumed. 
Values of $N$ greater than 1 have recently been found in severeal nearby galaxies by Momose et al. (2013) using CO(1-0) data to trace the molecular gas component. A different value of $N$ has been
obtained by the analysis of the THINGS sample of resolved galaxies at sub-kpc scales by Bigiel et al. (2008) and Leroy et al. (2013) 
whenever the molecular gas phase traced by the CO(2-1) transition 
alone is considered ($N$=1.0 $\pm$ 0.2). The physical interpretation of this relation in the framework of disc free-fall time, orbital timescale, 
cloud-cloud collisions, fixed giant molecular cloud efficiency, pressure of the different phases of the ISM, gravitational instability, galactic shear,
and formation of a cold phase has been extensively discussed in Leroy et al. (2008). \\

If $N$ $\simeq$ 1, as the THINGS survey indicates, then locally $\tau_{H_2}$ 
is constant. Overall our integrated results seem fairly consistent with those locally found by Leroy et al. (2008, 2013) on bright, nearby resolved galaxies. 
%The observed trend between $\tau_{H_2}$ and $SSFR$ (Fig. \ref{taucos} and \ref{tauvar}), is apparently in contraddiction with the results 
%of Leroy et al. (2008). 
We notice, however, that if we add to the late-type HRS the COLD GASS galaxies detected in
H$_2$, which mainly sample the lower end in $SSFR$ (grey dots in panel f), or the early-type HRS, the relation becomes flat. It seems thus 
that this relation is not universal\footnote{Caution should, however, be used in interpreting this evidence since the SFR of early-type galaxies is poorly constrained. 
Indeed, their UV emission can be due to evolved stars (O'Connell 1999; Boselli et al. 2005).}.
Our analysis indicates that a population of objects exists with high 
SSFRs ($SSFR$ $\gtrsim$ 5 10$^{-10}$ yr$^{-1}$) for which $\tau_{H_2}$ is smaller than in other galaxies (see Table \ref{Tabtaudata}).
These are mainly low-luminosity, metal-poor systems characterised by arelatively high star formation activity, probably undersampled in the
THINGS sample with the HERACLES CO data of Leroy et al. (2008, 2013).
The lack of resolved data for the gaseous component, however,
does not allow us to see whether, as claimed by the THINGS team, $\tau_{H_2}$ is constant as a function of the molecular gas density, or it changes as proposed by Momose et al. (2013).
At the same time, the lack of CO data for the HRS at different frequencies does not allow us to see whether, as claimed by Momose et al. (2013), the star 
formation efficiency changes when measured using different gas transitions. To conclude, the observed trend between  $\tau_{H_2}$ and $SSFR$ observed in the late-type HRS can be due
either to a change in the star formation efficiency $A$ or to the slope $N$ $\neq$ 1 of the Schmidt law in galaxies characterised by a different star formation history, but also to a 
variation in the $X_{CO}$ conversion factor that is larger than the one traced by the $X_{CO}$ vs. H-band luminosity relation of Boselli et al. (2002) assumed in this work (Leroy et al. 2013).

\subsection{Comparison with high-redshift galaxies}

Determining molecular and atomic gas scaling relations in high-redshift objects is still impossible because of the lack of statistically significant 
samples of galaxies. The most distant late-type galaxies observed and detected in H{\sc i} is at $z$ = 0.25 (Catinella et al. 2008).
The first direct observation of high-level CO transitions (Greve et al. 2005; Daddi et al. 2010; Genzel et al. 2010; 
Combes et al. 2013) or determination of molecular and atomic gas masses indirectly from the dust emission (Magdis et al. 2012a,b) have 
allowed the first determination of the molecular gas content of high-redshift objects. Combining these data with SFRs estimated using different techniques,
these authors determined star formation efficiencies or gas depletion timescales that can be compared directly to those deduced in this work.
A direct comparison of the different sets of data is also fair because these high-$z$ galaxies' gas consumption timescales have been determined by using integrated data
and are thus directly comparable to our own estimates. The main difference in the local and high-$z$ samples comes principally from the target selection.
While the HRS sample analysed in this work is K-band selected, and thus includes mainly normal star-forming galaxies, high-redshift objects are generally active star-forming galaxies.
Our analysis indicates that $\tau_{H_2}$ $\simeq$ 0.6$\pm$0.8 Gyr, only mildly dependent on the assumed $X_{CO}$ conversion factor, 
and is thus consistent with the mean value found by Genzel et al. (2010) for star-forming galaxies ($\tau_{H_2}$ = 0.5 Gyr at $z$ $\sim$ 2
and $\tau_{H_2}$ = 1.5 Gyr at $z$ $\sim$ 0), by Daddi et al. (2010) for BzK galaxies ($\tau_{H_2}$ = 0.3-0.8 Gyr), by Tacconi et al. (2013) for 
main sequence galaxies at $z$=1-3 ($\tau_{H_2}$ $\simeq$ 0.7 Gyr), but significantly larger than that of 
SMG and ULIRG ($\tau_{H_2}$ $\simeq$ 0.1 Gyr, Daddi et al. 2010; $\tau_{H_2}$ $\lesssim$ 0.1 Gyr, Combes et al. 2013).
We do not want to discuss here the possible origin of this systematic difference between the gas consumption timescale observed in star-forming vs. starburst galaxies
at high redshift. We just want to notice that our HRS data are consistent with a non-evolving star formation efficiency with $z$ in normal star-forming galaxies,
as claimed by Genzel et al. (2010).\\

\begin{figure*}
   \centering
   \includegraphics[width=14cm]{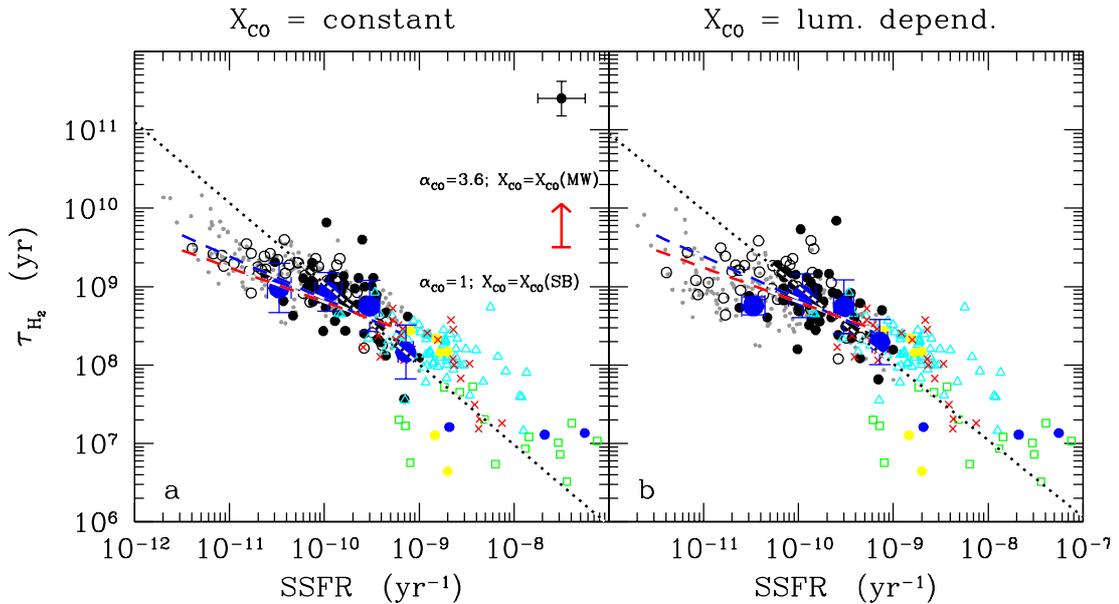}
   \caption{The relationship between molecular gas depletion timescale and the specific star formation rate 
   for HRS galaxies (black symbols), COLD GASS galaxies (grey symbols), and high-redshift objects, from Tacconi et al. (2008) (yellow filled dots), Daddi et al. (2009a,b) (blue filled dots),
   Genzel et al. (2010) (red crosses), Combes et al. (2013) (green open squares), and Tacconi et al. (2013) (cyan open triangles). The molecular gas content of all high-$z$ galaxies regardless of whether they 
   are star-forming or starburst galaxies, has been determined using a starburst conversion factor $X_{CO}$ = 0.64 10$^{20}$ cm$^{-2}$ (K km s$^{-1}$)$^{-1}$
   In the left panel the H$_2$ gas mass of HRS and COLD GASS galaxies is measured using a constant $X_{CO}$
   conversion factor, and in the right panel the luminosity-dependent $X_{CO}$ factor of Boselli et al. (2002). Both the local (HRS and COLD GASS)
   and the high-$z$ samples include only detected late-type galaxies. The big blue circles indicate the mean 
   values of $\tau_{H_2}$ in different bins of $SSFR$ determined from the HRS sample. The black dotted line shows the best fit determined for HRS galaxies, the coloured lines the best fits
   for the COLD GASS sample. The red vertical arrow indicate the systematic shift that all high-redshift galaxies would have if the molecular gas mass is determined assuming a Galactic
   conversion factor.
   }
   \label{tauz}%
   \end{figure*}

Figure \ref{tauz} shows the relationship between the molecular gas depletion timescale $\tau_{H_2}$ and the SSFR
for the HRS and COLD GASS galaxies and high-redshift star-forming and starburst objects taken from Tacconi et al. (2008), Daddi et al. (2009a, 2009b), Genzel et al. (2010),
and Combes et al. (2013). For all these high-$z$ galaxies, the molecular gas mass is estimated adopting a starburst conversion factor $\alpha_{CO}$ = 1 M$_{\odot}$ (K km s$^{-1}$ pc$^2$)$^{-1}$,
corresponding to $X_{CO}$ = 0.64 10$^{20}$ cm$^{-2}$ (K km s$^{-1}$)$^{-1}$, a value that is a factor of 3.6 lower than the constant value used in this work.
Figure \ref{tauz} shows that, under these assumptions on $X_{CO}$, high-redshift galaxies, regardless of whether they are star-forming or starburst galaxies, 
roughly follow and extend to lower molecular gas depletion timescales and
higher specific star formation rates the $\tau_{H_2}$ vs. $SSFR$ relation depicted by local objects. 

Clearly this result strongly depends on the assumption of a starburst 
conversion factor: the use of a Galactic value would increase $\tau_{H_2}$ by 0.56 dex (as shown in Fig. \ref{tauz}), thus changing the shape of the scaling
relation completely. Values of $X_{CO}$ close to the canonical Milky Way value have been estimated in the starburst galaxy M82 (Wild et al. 1992) and more recently 
in the Antennae by Schirm et al. (2013).
The correct value of the CO-to-H$_2$ conversion factor, which is one of the major source of uncertainty in local galaxies, becomes even more important
in high-redshift objects, where the physical conditions of the ISM might change more drastically than in the local universe. This topic, however, is beyond
the scope of the present work. We thus refer
the reader to dedicated publications (Daddi et al. 2010; Genzel et al. 2012, Combes et al. 2013). \\
We can, however, conclude that the analysis of Fig. \ref{tauz} does not show any
strong evidence of a variation of the gas consumption timescale with $z$, suggesting that the Schmidt law is invariant with cosmic time. The claimed variation 
in the star formation efficiency of galaxies, with two distinct regimes for normal and starburst objects, is also very weak since it is only based
on the assumption that the $X_{CO}$ conversion factor of starbursts is significantly and systematically different from that of normal, star-forming objects. 
The recent results of Momose et al. (2013) also indicate that the relationship between gas content and star formation can strongly depend on the density
of the molecular gas phase, as traced by different CO transitions. Any possible systematic variation in the gas depletion timescale between HRS (and COLD GASS)
galaxies and objects at high redshift, where the molecular gas content has been determined from high CO transitions, still needs to be confirmed.

\section{Conclusion}

The present work studied the properties of the cold gas component of the interstellar medium of the \textit{Herschel} Reference Survey, a complete volume-limited (15$\lesssim$ $D$ $\lesssim$ 25 Mpc), 
K-band-selected sample of galaxies spanning a wide range in morphological type and stellar mass. This work is based on public or newly obtained CO and H{\sc i} line measurements available for 
225 and 315 galaxies, respectively, out of the 322 objects composing the sample.
The analysis has shown that the molecular gas mass distribution of the HRS is very different from the CO luminosity distribution determined from the FCRAO sample, a far infrared selected 
sample of nearby objects, by  Keres et al. (2003) and transformed into a molecular gas mass distribution by Obreschkow \& Rawlings (2009). 
This result clearly shows how molecular gas mass distributions
derived from pointed observations differ from each other, and thus cannot be taken as representative of real mass functions determined from CO blind surveys
that are unfortunately still lacking.\\

We have extended the main scaling relations analysed by Saintonge et al. (2011a,b) from the COLD GASS sample, limited in the stellar mass range 
10$^{10}$ $\lesssim$ $M_{star}$ $\lesssim$ 2 10$^{11}$ M$_{\odot}$, down to $M_{star}$ $\simeq$ 10$^{9}$ M$_{\odot}$. We also estimated how these relations are sensitive to the use of a constant or
a luminosity-dependent $X_{CO}$ conversion factor. These scaling relations trace the variations in the total and molecular gas-to-stellar mass ratio, of the atomic-to-molecular gas 
ratio, and of the total and molecular gas depletion timescale as a function of morphological type, total stellar mass, total stellar mass surface density, SSFR, and metallicity of the HRS
galaxies. The analysis has shown that\\
1) The $M(gas)/M_{star}$ ratio increases with morphological type, while $M(H_2)/M_{star}$ and $M(H_2)/M(\mathrm{\hi})$ are fairly constant along the spiral sequence. In late-type galaxies the molecular
hydrogen content is $\simeq$ 25-30 \% ~ of the atomic gas.
Early types have significantly lower fractions of molecular and total gas than the late-type systems of similar stellar mass. \\
2) The relations concerning $M(gas)/M_{star}$ are always steeper and less dispersed than those with $M(H_2)/M_{star}$.\\
3) The slope of the relations strongly depends on the adopted conversion factor. The only relations that remain statistically significant using both a constant and a luminosity-dependent $X_{CO}$,
and thus can be considered robust vs. the very uncertain value of the conversion factor, are all those concerning $M(gas)/M_{star}$. On the contrary, only the observed increase
in $M(H_2)/M_{star}$ on $SSFR$ and of $M(H_2)/M(\mathrm{\hi})$ vs. $\mu_{star}$ and 12+log(O/H) are statistically significant regardless of $X_{CO}$. Similarly, 
the only statistically significant scaling relations concerning the gas consumption timescale are those showing a decrease in $\tau_{gas}$ with $\mu_{star}$ and 12+log(O/H) 
and that of $\tau_{H_2}$ with the SSFR.\\
4) We also observed a change in slope in many relations when the HRS galaxies are combined with the COLD GASS sample, this last dominated by massive, quiescent objects. Some of 
these relations are thus non linear.\\
5) The relationship between the molecular-to-atomic gas ratio and the metallicity is predicted by recent physical models of the ISM.\\
6) The star formation efficiency of normal, star-forming galaxies is invariant with cosmic time. The claimed variation in the star formation efficiency
from star-forming and starburst galaxies is only based on the assumption that the $X_{CO}$ conversion factor is systematically different in these
two regimes. 

The results obtained so far give new strong observational constraints to models of galaxy formation and evolution. They can be compared
to the most recent semi-analytic models, which are now able to reproduce the evolution of the different gaseous phases, to understand the complex process that gave birth
to local galaxies.

\begin{acknowledgements}

A. B. thanks the ESO visiting program committee for inviting him
to the Garching headquarters for a two-months stay. We are also grateful to M. Fossati
for his help during the preparation of the CO dataset. We thank the anonymous referee for precious comments and suggestions that helped improve the quality of the manuscript.
The research leading to these results has received funding from the European Community's Seventh 
Framework Programme (/FP7/2007-2013/) under grant agreement No 229517. 
B.C. is the recipient of an Australian Research Council Future Fellowship (FT120100660).
This research made use of the 
NASA/IPAC Extragalactic Database (NED), 
which is operated by the Jet Propulsion Laboratory, California Institute of 
Technology, under contract with the National Aeronautics and Space Administration
and of the GOLDMine database (http://goldmine.mib.infn.it/).
IRAF is distributed by the National Optical Astronomy Observatory, 
which is operated by the Association of Universities for Research in Astronomy 
(AURA) under cooperative agreement with the National Science Foundation.

\end{acknowledgements}

%\clearpage
\begin{table*}
\caption{The cold gas content of unperturbed galaxies of different morphological types (big red filled triangles and big blue filled dots in Figure 4)}
\label{Tabtype}
{%\tiny
\[
\begin{tabular}{ccccccccccccccc}
\hline
\hline
\noalign{\smallskip}
\hline
Gas	               	& $X_{CO}$& E (6)		&S0 (9)			    & Sa (8)		    & Sab (4)		    & Sb (10)		    & Sbc (17)  	    & Sc (25)		    & Scd (14)      & Sd (7)	    & Sdm (5)	    & Im (3) \\
\hline
log$M(gas)/M_{star}$    & const	 & -2.07$\pm$0.50	& -1.27$\pm$0.40	    & -0.33$\pm$0.45	    & -0.80$\pm$0.16	    & -0.33$\pm$0.24	    & -0.33$\pm$0.23	    & -0.20$\pm$0.26	    & 0.00$\pm$0.37 & 0.10$\pm$0.31 & 0.33$\pm$0.21 & 0.08$\pm$0.39\\
log$M(gas)/M_{star}$    &lum.dep.& -2.18$\pm$0.56	& -1.32$\pm$0.44	    & -0.30$\pm$0.49	    & -0.86$\pm$0.13	    & -0.33$\pm$0.24	    & -0.37$\pm$0.25	    & -0.23$\pm$0.31	    & 0.00$\pm$0.36 & 0.12$\pm$0.30 & 0.37$\pm$0.24 & 0.12$\pm$0.36\\
log$M(H_2)/M_{star}$    & const	 & -2.59$\pm$0.47	& -2.10$\pm$0.28	    & -0.99$\pm$0.54	    & -1.63$\pm$0.48	    & -1.07$\pm$0.33	    & -1.04$\pm$0.40	    & -0.92$\pm$0.23	    & -1.10$\pm$0.26& -0.94$\pm$0.26& -0.68$\pm$0.45& -0.79$\pm$0.11\\
log$M(H_2)/M_{star}$    &lum.dep.& -2.84$\pm$0.60	& -2.25$\pm$0.28	    & -0.95$\pm$0.64	    & -1.93$\pm$0.40	    & -1.11$\pm$0.43	    & -1.17$\pm$0.40	    & -0.98$\pm$0.21	    & -1.04$\pm$0.21& -0.85$\pm$0.29& -0.54$\pm$0.54& -0.63$\pm$0.17\\
log$M(H_2)/M(\mathrm{\hi})$	& const	 &  0.03$\pm$0.74	& -0.50$\pm$0.65	    & -0.35$\pm$0.42	    & -0.60$\pm$0.46	    & -0.49$\pm$0.26	    & -0.53$\pm$0.49	    & -0.42$\pm$0.48	    & -0.90$\pm$0.52& -0.85$\pm$0.33& -0.82$\pm$0.36& -0.62$\pm$0.60\\  
log$M(H_2)/M(\mathrm{\hi})$	&lum.dep.& -0.23$\pm$0.80	& -0.66$\pm$0.63	    & -0.31$\pm$0.48	    & -0.89$\pm$0.39	    & -0.53$\pm$0.35	    & -0.40$\pm$0.52	    & -0.48$\pm$0.40	    & -0.84$\pm$0.46& -0.77$\pm$0.38& -0.68$\pm$0.45& -0.46$\pm$0.61\\  
\noalign{\smallskip}
\hline
\end{tabular}
\]
%\scriptsize
Note: mean values and standard deviations are determined considering only H{\sc i} detected galaxies and CO upper limits as detections; for late-type galaxies, means are 
determined using only unperturbed galaxies with $\mathrm{\hi}-def$ $<$ 0.4. The number in parenthesis in each morphological class gives the number of objects used to determine the mean values. }
\end{table*}

\clearpage

\begin{table*}
\caption{Coefficients of the scaling relations: y = ax + b (dotted lines in Figs. 5 and 6)}
\label{Tabscalingfit}
{%\scriptsize
\[
\begin{tabular}{ccccccc}
\hline
\hline
\noalign{\smallskip}
\hline
y			&x			& $X_{CO}$       & a     & b     & $\rho$ &$\sigma$ \\
\hline
log$M(gas)/M_{star}$	& log$M_{star}$		& constant       &-0.69  & 6.63  &-0.69  & 0.22 \\ % a
			&			& lum.dep.       &-0.74  & 7.03  &-0.77  & 0.20 \\ % a			
			& log$\mu_{star}$	& constant       &-0.78  & 6.00  &-0.64  & 0.22 \\ % a
			&			& lum.dep.       &-0.83  & 6.41  &-0.66  & 0.23 \\ % a			 
			& log$SSFR$		& constant       & 0.94  & 9.00  & 0.70  & 0.17 \\ % a
			&			& lum.dep.       & 1.00  & 9.60  & 0.67  & 0.19 \\ % a
			& 12+log(O/H)		& constant       &-2.42  &20.72  &-0.66  & 0.09 \\ % a
			&			& lum.dep.       &-2.63  &22.51  &-0.70  & 0.09 \\ % a
\hline
log$M(H_2)/M_{star}$	& log$M_{star}$		& constant       &-0.93  & 8.12  &-0.03  & 0.38 \\ % a
			&			& lum.dep.       &-0.76  & 6.40  &-0.52  & 0.28 \\ % a			 
			& log$\mu_{star}$	& constant       & 0.97  &-8.63  & 0.14  & 0.36 \\ % a
			&			& lum.dep.       &-0.92  & 6.29  &-0.12  & 0.35 \\ % a			 
			& log$SSFR$		& constant       & 0.94  & 8.19  & 0.47  & 0.25 \\ % a
			&			& lum.dep.       & 1.01  & 8.92  & 0.54  & 0.24 \\ % a
			& 12+log(O/H)		& constant       &-1.24  & 9.74  & 0.01  & 0.18 \\ % a
			&			& lum.dep.       &-1.90  &15.50  &-0.30  & 0.14 \\ % a
\hline
log$M(H_2)/M(\mathrm{\hi})$& log$M_{star}$	& constant       & 1.01  &-10.46 & 0.58  & 0.33 \\ % a
			&			& lum.dep.       & 0.95  &-9.86  & 0.27  & 0.40 \\ % a
			& log$\mu_{star}$	& constant       & 1.13  &-9.46  & 0.63  & 0.29 \\ % a
			&			& lum.dep.       & 1.01  &-8.52  & 0.53  & 0.31 \\ % a
			& log$SSFR$		& constant       &-1.20  &-12.29 &-0.19  & 0.36 \\ % a
			&			& lum.dep.       &-1.09  &-11.19 &-0.11  & 0.36 \\ % a
			& 12+log(O/H)		& constant       & 3.40  &-29.86 & 0.59  & 0.13 \\ % a
			&			& lum.dep.       & 2.66  &-23.51 & 0.47  & 0.15 \\ % a
\noalign{\smallskip}
\hline
\end{tabular}
\]
Notes: bisector fit. The sample used to fit the relation is composed of all the H{\sc i} and CO detected late-type galaxies with $\mathrm{\hi}-def$ $<$ 0.4. \\
$\rho$ gives the Spearman correlation coefficient, $\sigma$ the dispersion in the relations.\\
}
\end{table*}

%\clearpage
\begin{table*}
\caption{Average scaling relations (big blue filled dots Figs. 5 and 6)}
\label{Tabscalingdata}
{%\scriptsize
\[
\begin{tabular}{cccccc}
\hline
\hline
\noalign{\smallskip}
%\hline
\multicolumn{3}{c}{}&\multicolumn{1}{c}{$X_{CO}$ constant}&\multicolumn{1}{c}{$X_{CO}$ lum.dep.}&\multicolumn{1}{c}{}\\
x		&	y		&  $<x>$ &	 $<y>$   & $<y>$	 & N  \\
\hline
log$M_{star}$	& log$M(gas)/M_{star}$	& 9.15	 & 0.27$\pm$0.26 & 0.30$\pm$0.26 & 10 \\ % a
		&			& 9.49	 &-0.07$\pm$0.28 &-0.06$\pm$0.27 & 30 \\ % a
		&			& 10.00	 &-0.33$\pm$0.23 &-0.37$\pm$0.23 & 23 \\ % a
		&			& 10.52	 &-0.37$\pm$0.18 &-0.48$\pm$0.18 & 11 \\ % a
		& log$M(H_2)/M_{star}$	& 9.15	 &-0.81$\pm$0.40 &-0.66$\pm$0.40 & 10 \\ % a
		&			& 9.49	 &-1.03$\pm$0.20 &-0.95$\pm$0.21 & 30 \\ % a
		&			& 10.00	 &-0.96$\pm$0.36 &-1.07$\pm$0.36 & 23 \\ % a
		&			& 10.52	 &-0.85$\pm$0.26 &-1.12$\pm$0.26 & 11 \\ % a
		& log$M(H_2)/M(\mathrm{\hi})$	& 9.15	 &-0.90$\pm$0.49 &-0.74$\pm$0.50 & 10 \\ % a
		&			& 9.49	 &-0.75$\pm$0.35 &-0.67$\pm$0.36 & 30 \\ % a
		&			& 10.00	 &-0.27$\pm$0.49 &-0.38$\pm$0.49 & 23 \\ % a
		&			& 10.52	 &-0.09$\pm$0.27 &-0.36$\pm$0.27 & 11 \\ % a
\hline					
log$\mu_{star}$ & log$M(gas)/M_{star}$	& 7.39 	 & 0.10$\pm$0.23 & 0.11$\pm$0.24 & 17 \\ % a
		&			& 7.88 	 &-0.15$\pm$0.28 &-0.16$\pm$0.29 & 30 \\ % a
		&			& 8.34 	 &-0.39$\pm$0.26 &-0.44$\pm$0.26 & 24 \\ % a
		&			& 8.77 	 &-0.46$\pm$0.10 &-0.52$\pm$0.15 &  3 \\ % a
		& log$M(H_2)/M_{star}$	& 7.39	 &-1.08$\pm$0.19 &-0.99$\pm$0.26 & 17 \\ % a
		&			& 7.88	 &-0.92$\pm$0.25 &-0.95$\pm$0.31 & 30 \\ % a
		&			& 8.34	 &-0.96$\pm$0.36 &-1.08$\pm$0.38 & 24 \\ % a
		&			& 8.77	 &-0.89$\pm$0.03 &-1.04$\pm$0.22 &  3 \\ % a
		& log$M(H_2)/M(\mathrm{\hi})$	& 7.39	 &-1.01$\pm$0.23 &-0.92$\pm$0.24 & 17 \\ % a
		&			& 7.88	 &-0.53$\pm$0.31 &-0.56$\pm$0.29 & 30 \\ % a
		&			& 8.34	 &-0.19$\pm$0.47 &-0.31$\pm$0.46 & 24 \\ % a
		&			& 8.77	 &-0.01$\pm$0.24 &-0.16$\pm$0.33 &  3 \\ % a
\hline
log$SSFR$	& log$M(gas)/M_{star}$	&-10.48	 &-0.61$\pm$0.20 &-0.65$\pm$0.23 &  4 \\ % a
		&			& -9.96  &-0.29$\pm$0.22 &-0.31$\pm$0.24 & 37 \\ % a
		&			& -9.52  &-0.01$\pm$0.28 &-0.03$\pm$0.31 & 22 \\ % a
		&			& -9.14  & 0.34$\pm$0.17 & 0.35$\pm$0.16 &  6 \\ % a
		& log$M(H_2)/M_{star}$	&-10.48	 &-1.49$\pm$0.22 &-1.72$\pm$0.12 &  4 \\ % a
		&			& -9.96  &-1.02$\pm$0.24 &-1.07$\pm$0.26 & 37 \\ % a
		&			& -9.52  &-0.77$\pm$0.29 &-0.78$\pm$0.31 & 22 \\ % a
		&			& -9.14  &-0.98$\pm$0.35 &-0.55$\pm$0.29 &  6 \\ % a
		& log$M(H_2)/M(\mathrm{\hi})$	&-10.48	 &-0.66$\pm$0.43 &-0.89$\pm$0.32 &  4 \\ % a
		&			& -9.96  &-0.43$\pm$0.46 &-0.48$\pm$0.43 & 37 \\ % a
		&			& -9.52  &-0.47$\pm$0.50 &-0.48$\pm$0.42 & 22 \\ % a
		&			& -9.14  &-1.16$\pm$0.42 &-1.04$\pm$0.37 &  6 \\ % a

\hline
12+logO/H	& log$M(gas)/M_{star}$	& 8.34	 & 0.33$\pm$0.24 & 0.38$\pm$0.28 &  3 \\ % a
		&			& 8.52   & 0.10$\pm$0.24 & 0.12$\pm$0.24 & 23 \\ % a
		&			& 8.70   &-0.29$\pm$0.21 &-0.33$\pm$0.22 & 36 \\ % a
		& log$M(H-2)/M_{star}$	& 8.34	 &-0.64$\pm$0.55 &-0.46$\pm$0.60 &  3 \\ % a
		&			& 8.52   &-0.96$\pm$0.24 &-0.88$\pm$0.27 & 23 \\ % a
		&			& 8.70   &-0.91$\pm$0.26 &-0.99$\pm$0.26 & 36 \\ % a
		& log$M(H_2)/M(\mathrm{\hi})$	& 8.34	 &-0.76$\pm$0.48 &-0.58$\pm$0.54 &  3 \\ % a
		&			& 8.52   &-0.88$\pm$0.32 &-0.80$\pm$0.32 & 23 \\ % a
		&			& 8.70   &-0.27$\pm$0.40 &-0.36$\pm$0.37 & 36 \\ % a
\noalign{\smallskip}
\hline
\end{tabular}
\]
Note: mean values and standard deviations for the scaling relations determined using only H{\sc i} and CO detected, unperturbed ($\mathrm{\hi}-def$ $<$ 0.4) late-type galaxies }
\end{table*}

\begin{table*}
\caption{Coefficients of the gas depletion time relations: log$\tau$ = ax + b (dotted lines in Figs. 7 and 8)}
\label{Tabtaufit}
{%\scriptsize
\[
\begin{tabular}{ccccccc}
\hline
\hline
\noalign{\smallskip}
\hline
y		&x			& $X_{CO}$       & a     & b     & $\rho$ & $\sigma$ \\
\hline
log$\tau_{gas}$	& log$M_{star}$		& constant       &-0.85  &17.97  &-0.12  & 0.34 \\ % a
		&			& lum.dep.       &-0.73  &16.75  &-0.29  & 0.30 \\ % a
		& log$\mu_{star}$	& constant       &-0.66  &14.82  &-0.39  & 0.24 \\ % a
		&			& lum.dep.       &-0.68  &14.95  &-0.43  & 0.24 \\ % a
		& log$SSFR$		& constant       &-0.77  & 2.06  &-0.45  & 0.21 \\ % a
		&			& lum.dep.       &-0.84  & 1.41  &-0.41  & 0.24 \\ % a
		& 12+log(O/H)		& constant       &-1.62  &23.54  &-0.32  & 0.13 \\ % a
		&			& lum.dep.       &-1.84  &25.41  &-0.37  & 0.13 \\ % a
\hline
log$\tau_{H_2}$	& log$M_{star}$		& constant       & 0.79  & 1.03  & 0.52  & 0.30 \\ % a
		&			& lum.dep.       & 0.99  &-0.89  &-0.01  & 0.41 \\ % a
		& log$\mu_{star}$	& constant       & 0.87  & 1.88  & 0.36  & 0.31 \\ % a
		&			& lum.dep.       & 0.93  & 1.42  & 0.15  & 0.36 \\ % a
		& log$SSFR$		& constant       &-1.03  &-1.25  &-0.57  & 0.22 \\ % a
		&			& lum.dep.       &-0.98  &-0.78  &-0.52  & 0.25 \\ % a
		& 12+log(O/H)		& constant       & 1.89  &-7.47  & 0.39  & 0.16 \\ % a
		&			& lum.dep.       & 1.18  &-1.34  & 0.14  & 0.22 \\ % a
\noalign{\smallskip}
\hline
\end{tabular}
\]
Notes: bisector fit. The sample used to fit the relation is composed of all the H{\sc i}- and CO-detected late-type galaxies with $\mathrm{\hi}-def$ $<$ 0.4. \\
$\rho$ gives the Spearman correlation coefficient, $\sigma$ the dispersion in the relations.\\
}
\end{table*}

%\clearpage

\begin{table*}
\caption{Average gas depletion time relations (big blue filled dots in Figs. 7 and 8)}
\label{Tabtaudata}
{%\scriptsize
\[
\begin{tabular}{cccccc}
\hline
\hline
\noalign{\smallskip}
%\hline
\multicolumn{3}{c}{}&\multicolumn{1}{c}{$X_{CO}$ constant}&\multicolumn{1}{c}{$X_{CO}$ lum.dep.}&\multicolumn{1}{c}{}\\
x		&	y		&  $<x>$ &	 $<y>$   & $<y>$	 & N  \\
\hline
log$M_{star}$	& log$\tau_{gas}$	& 9.16	 & 9.61$\pm$0.26 & 9.64$\pm$0.28 &  9 \\ % a
		&			& 9.50	 & 9.65$\pm$0.24 & 9.66$\pm$0.24 & 27 \\ % a
		&			& 10.01	 & 9.59$\pm$0.25 & 9.56$\pm$0.27 & 22 \\ % a
		&			& 10.49	 & 9.47$\pm$0.17 & 9.37$\pm$0.20 &  9 \\ % a
		& log $\tau_{H_2}$	& 9.16	 & 8.53$\pm$0.58 & 8.68$\pm$0.60 &  9 \\ % a
		&			& 9.50	 & 8.69$\pm$0.26 & 8.77$\pm$0.31 & 27 \\ % a
		&			& 10.01	 & 8.97$\pm$0.27 & 8.86$\pm$0.30 & 22 \\ % a
		&			& 10.49	 & 8.99$\pm$0.13 & 8.73$\pm$0.14 &  9 \\ % a
\hline
log$\mu_{star}$	& log$\tau_{gas}$	& 7.39	 & 9.75$\pm$0.22 & 9.76$\pm$0.23 & 15 \\ % a
		&			& 7.87	 & 9.58$\pm$0.21 & 9.57$\pm$0.22 & 27 \\ % a
		&			& 8.33	 & 9.57$\pm$0.24 & 9.52$\pm$0.25 & 22 \\ % a
		&			& 8.77	 & 9.33$\pm$0.22 & 9.27$\pm$0.26 &  3 \\ % a
		& log $\tau_{H_2}$	& 7.39   & 8.57$\pm$0.33 & 8.64$\pm$0.36 & 15 \\ % a
		&			& 7.87   & 8.82$\pm$0.21 & 8.79$\pm$0.23 & 27 \\ % a
		&			& 8.33   & 8.98$\pm$0.32 & 8.86$\pm$0.29 & 22 \\ % a
		&			& 8.77   & 8.89$\pm$0.11 & 8.75$\pm$0.26 &  3 \\ % a
\hline
log$SSFR$	& log$\tau_{gas}$	&-10.48	 & 9.87$\pm$0.18 & 9.82$\pm$0.18 &  4 \\ % a
		&			& -9.96	 & 9.66$\pm$0.23 & 9.65$\pm$0.25 & 37 \\ % a
		&			& -9.52	 & 9.51$\pm$0.24 & 9.49$\pm$0.28 & 22 \\ % a
		&			& -9.14	 & 9.48$\pm$0.13 & 9.49$\pm$0.13 &  6 \\ % a
		& log $\tau_{H_2}$	&-10.48  & 8.98$\pm$0.31 & 8.75$\pm$0.12 &  4 \\ % a
		&			& -9.96  & 8.93$\pm$0.25 & 8.89$\pm$0.28 & 37 \\ % a
		&			& -9.52  & 8.75$\pm$0.33 & 8.74$\pm$0.34 & 22 \\ % a
		&			& -9.14  & 8.17$\pm$0.34 & 8.29$\pm$0.29 &  6 \\ % a
\hline
12+logO/H	& log$\tau_{gas}$	& 8.52	 & 9.69$\pm$0.24 & 9.70$\pm$0.24 & 24 \\ % a
		&			& 8.70	 & 9.51$\pm$0.21 & 9.48$\pm$0.23 & 34 \\ % a
		& log $\tau_{H_2}$	& 8.52   & 8.58$\pm$0.34 & 8.66$\pm$0.35 & 22 \\ % a
		&			& 8.70   & 8.91$\pm$0.24 & 8.82$\pm$0.25 & 34 \\ % a
\noalign{\smallskip}
\hline
\end{tabular}
\]
Note: mean values and standard deviations for the scaling relations determined using only H{\sc i}- and CO-detected, unperturbed ($\mathrm{\hi}-def$ $<$ 0.4) late-type galaxies }
\end{table*}

\end{document}